\documentclass[usenatbib]{mnras}
\usepackage[fleqn]{amsmath}
\usepackage{amssymb}
\usepackage{morefloats}
\usepackage{graphicx}
\usepackage{array}
\usepackage{siunitx}
\usepackage{subfig}
\usepackage{lpic}
\usepackage{textgreek}
\usepackage{wrapfig}
\usepackage[normalem]{ulem}

\newcolumntype{L}[1]{>{\raggedright\let\newline\\\arraybackslash\hspace{0pt}}p{#1}}
\newcolumntype{C}[1]{>{\centering\let\newline\\\arraybackslash\hspace{0pt}}p{#1}}
\newcolumntype{R}[1]{>{\raggedleft\let\newline\\\arraybackslash\hspace{0pt}}p{#1}}

\title[Dust formation in SN 2005ip]{A decade of ejecta dust formation in the Type IIn SN~2005ip }

\author[A. Bevan et al.]{{A. Bevan$^{1}$\thanks{antonia.bevan.12@ucl.ac.uk},
R. Wesson$^{1}$, 
M. J. Barlow$^{1}$, 
I. De Looze$^{1}$,}
J. E. Andrews$^{2}$, 
G. C. Clayton$^{3}$, 
\newauthor{K. Krafton$^{3}$, 
M. Matsuura$^{4}$ and 
D. Milisavljevic$^{5}$}  \\
$^{1}$Department of Physics and Astronomy, University College London,Gower Street, London WC1E 6BT, UK \\
$^{2}$Steward Observatory, University of Arizona, 933 North Cherry Avenue, Tucson, AZ 85721, USA\\
$^{3}$Department of Physics \& Astronomy, Louisiana State University, Baton Rouge, LA 70803, USA\\
$^{4}$School of Physics \& Astronomy, Cardiff University, The Parade, Cardiff, CF24 3AA, UK \\
$^{5}$Department of Physics and Astronomy, Purdue University, 525 Northwestern Ave., West Lafayette, IN 47907, USA}

\begin{document}

\pagerange{\pageref{firstpage}--\pageref{lastpage}} \pubyear{2019}

\maketitle

\label{firstpage}

\begin{abstract}

\noindent In order to understand the contribution of core-collapse supernovae to the dust budget of the early universe, it is important to understand not only the mass of dust that can form in core-collapse supernovae but also the location and rate of dust formation. SN~2005ip is of particular interest since dust has been inferred to have formed in both the ejecta and the post-shock region behind the radiative reverse shock. We have collated eight optical archival spectra that span the lifetime of SN~2005ip and we additionally present a new X-shooter optical--near-IR spectrum of SN~2005ip at 4075\,d post-discovery. Using the Monte Carlo line transfer code {\sc damocles}, we have modelled the blueshifted broad and intermediate width  H$\alpha$, H$\beta$ and He~{\sc i} lines from 48\,d to 4075\,d post-discovery using an ejecta dust model. We find that dust in the ejecta can account for the asymmetries observed in the broad and intermediate width H$\alpha$, H$\beta$ and He~{\sc i} line profiles at all epochs and that it is not necessary to invoke post-shock dust formation to explain the blueshifting observed in the intermediate width post-shock lines. Using a Bayesian approach, we have determined the evolution of the ejecta dust mass in SN 2005ip over 10 years presuming an ejecta dust model, with an increasing dust mass from $\sim 10^{-8}$~M$_{\odot}$ at 48\,d to a current dust mass of $\sim$0.1~M$_{\odot}$.

\end{abstract}


\section{Introduction}

The source of the large masses of dust observed in some very early Universe galaxies at redshifts $z>6$ has been much debated \citep{Omont2001,Bertoldi2003,Watson2015,Laporte2017}. 
Theoretical models predict that core-collapse supernovae (CCSNe) are capable of producing $>0.1$M$_{\odot}$ of ejecta-condensed dust.  \citep{Todini2001,Nozawa2003,Sarangi2015}. 
Generally, dust mass estimates in CCSNe have been made by fitting their infrared (IR) spectral energy distributions (SEDs), but, until the advent of {\em Herschel}, a dearth of observations at longer, far-IR wavelengths had led to the detection of only warm and hot supernova dust up to a maximum dust mass of $\sim10^{-3}$\,M$_{\odot}$ a few years after outburst \citep{Sugerman2006,Meikle2007,Andrews2010,Fabbri2011,Gomez2013,Gall2014}.  
Following the 2009\,--\,2013 {\em Herschel} mission, large masses ($>0.1$\,M$_{\odot}$) of cold ($<<100$\,K) dust in a few nearby CCSNe and remnants have been discovered \citep{Matsuura2011,Gomez2012, Indebetouw2014,Matsuura2015,Owen2015,deLooze2017,Temim2017,Rho2018}. Estimates of the dust masses that have formed in these objects are consistent with the estimated $\sim$0.1\,--\,1.0\,M$_{\odot}$ of dust per CCSN required to account for the dust observed in the early universe \citep{Morgan2003,Dwek2007}. While a few CCSNe now have late-time dust mass estimates, only one (SN~1987A) has had the evolution of dust in its ejecta observationally estimated at multiple epochs \citep{Wesson2015,Bevan2016}. The inferred gradual growth of the ejecta dust mass in SN~1987A over $>10$\,yr is in contrast to dust nucleation models which predict very rapid increases in dust mass over much shorter timescales \mbox{\citep[$<$1500\,d;][]{Sarangi2015,Sluder2018}}. A better understanding of dust formation and destruction in CCSNe is required in order to address the dust budget problem in the early Universe.  

Unfortunately, there is likely to be a long wait for a far-IR telescope that is capable of detecting cold dust in extragalactic CCSNe. However, SN dust formation has additional observational signatures besides far-IR emission that allow it to be traced. The presence of dust in CCSNe can induce persistent asymmetries in their optical and NIR line profiles due to greater attenuation of redshifted radiation which must traverse the dusty interior. Any dust interior to or colocated with the source of the line emission can induce a characteristic red-blue asymmetry where the peak flux and flux bias become shifted towards the blue. This was first noted for SN~1987A by \citet{Lucy1989}. \citet{Bevan2016} and \citet{Bevan2017} exploited this signature by using the radiative transfer code {\sc damocles} to model the extent and shape of these line asymmetries, allowing them to determine the ejecta-condensed dust mass at late times (700 -- 10000\,d) for a number of CCSNe.

Dust formation has been associated with a number of Type IIn supernovae \citep[e.g.][]{Mauerhan2012,Stritzinger2012,Gall2014}. SN 2005ip is  one such Type IIn SN that is of particular interest for a number of reasons. Its ongoing interaction with dense circumstellar material (CSM) ejected by the progenitor star prior to its explosion has caused it to remain bright at optical and near-IR wavelengths at its distance of 30\,Mpc for years after its outburst  \citep{deVaucouleurs1991,Stritzinger2012,Katsuda2014,Smith2017}.
It has been consistently observed both photometrically and spectroscopically, allowing the evolution of its line profiles to be studied. Analyses of these lines have already yielded significant insights into the physics of the object and can allow the dust mass formation rate to be ascertained  \citep{Smith2009,Stritzinger2012,Smith2017}.  

Based on the evolving blueshifted asymmetries in the optical and near-infrared (IR) line profiles and a persistent plateau in the IR light curve, dust has been deduced to have formed in  SN~2005ip in both the post-shock region that arises between the forward and reverse shocks and in the ejecta \citep{Fox2009,Smith2009,Fox2010,Stritzinger2012,Smith2017}. Dust mass estimates for SN~2005ip to date have been based on fits to near-IR and mid-IR ($<15${\si{\um}}) observations \citep{Fox2010,Stritzinger2012}.

Asymmetries that develop in the broad H$\alpha$ line profile at early times ($t<200$\,d) have been attributed to dust formation in the ejecta of SN~2005ip \citep{Smith2009}.  At later times ($t>400$\,d), the appearance of blueshifted He\,{\sc i} and Balmer series line profiles in the intermediate width emission lines from the post-shock gas have been attributed to dust formation in the post-shock region \citep{Smith2009,Smith2017}.  We consider here the possibility that continued dust formation in the ejecta of SN~2005ip could be the cause of the blueshifting observed in these late-time intermediate width line profiles.

In this paper, we collate optical spectra of SN~2005ip spanning an eight year period from 48\,d to 3099\,d post-discovery. We additionally present a new, very late-time optical\,--\,near-IR spectrum of SN2005ip obtained with X-shooter in 2016 (4075\,d post-discovery). We have used the {\sc damocles} line transfer code to model the evolution of the broad and intermediate width line profiles using a Bayesian methodology to explore and constrain the parameter space  \citep[][]{Bevan2016,damocles}\footnote[1]{http://ascl.net/1807.023}. Different geometries were adopted for the early-time broad line models and later-time intermediate width lines models, with the radiation emanating from the ejecta and post-shock regions respectively. We adopted a model where dust was located entirely within the ejecta in order to determine whether dust formation in the ejecta of SN~2005ip could account for the observed behaviour of the line profiles at all epochs. Based on these models, we have determined the condensed dust mass in SN~2005ip at a range of ages and deduced the dust formation rate over time.

In Section \ref{scn_05ip_spectra}, we present the new X-shooter spectrum of SN~2005ip 11\,yr after discovery and discuss the archival data that was collated for our study. We present the details of the codes and statistical methods used to constrain the parameter space in Section \ref{scn_model_setup} before presenting the models and their results in Section \ref{scn_all_model_results}. We discuss the implications and context of these results in Section \ref{scn_discussion} and conclude in Section \ref{scn_conclusions}.

\section{Observations of SN~2005ip}
\label{scn_05ip_spectra}

\begin{figure*}
	\includegraphics[clip = true, trim = 15 10 10 10, width = \linewidth]{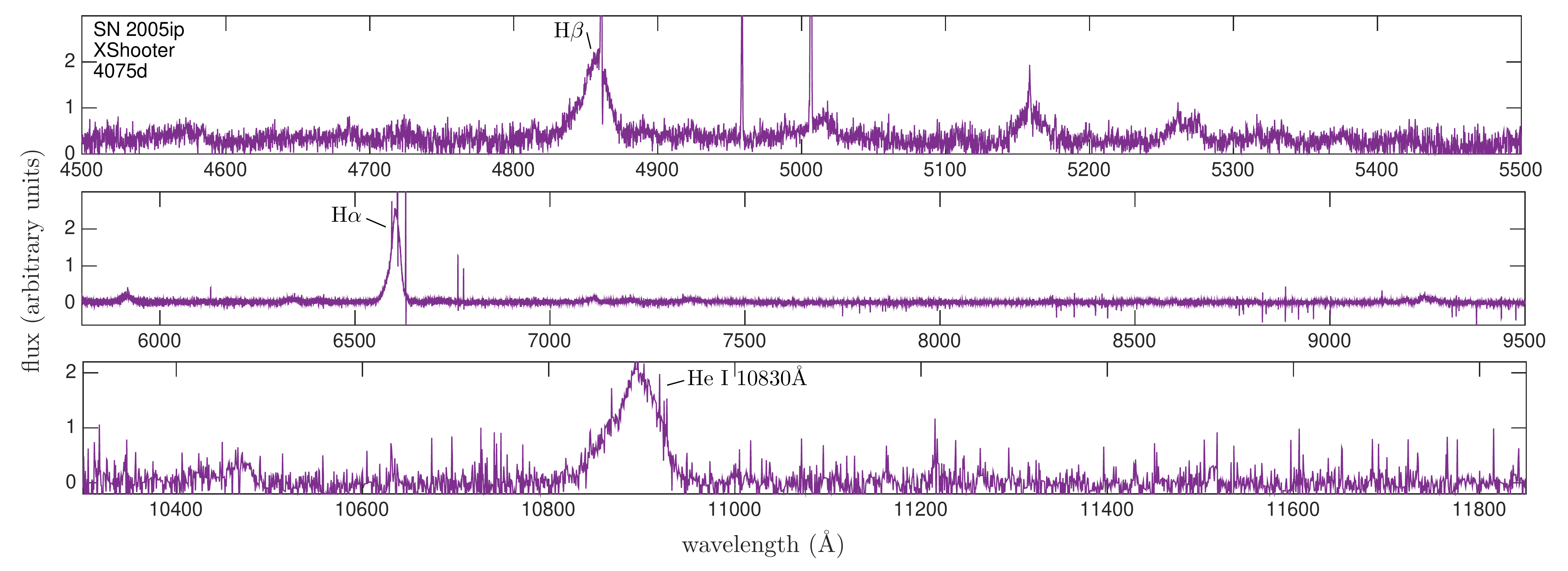}
	\caption{The UVB, visible and near-IR spectrum of SN~2005ip taken with X-shooter on the VLT at 4075~d post-discovery.}
	\label{fig_X-shooter_spectrum}
\end{figure*}

SN~2005ip is a Type IIn CCSN located  approximately 30~Mpc away in NGC 2906 at a redshift of z=0.00714 \citep{deVaucouleurs1991,Smith2009}.  It was discovered on 2005 November 5.2 \citep{Boles2005} and originally classified as a Type II supernova by \citet{Modjaz2005}. Further inspection of the early-time spectra revealed narrow line emission which led to its final classification as a Type IIn  \citep{Smith2009}. Based on  blackbody fits to the dereddened continuum and the maximum shock velocity as inferred from line profiles, \citet{Smith2009} estimated the explosion date to be 8--10 days prior to discovery. All ages given in this paper are given relative to the date of discovery. This likely only impacts the dust masses inferred at the very earliest times.

SN~2005ip has been well-observed since its outburst. \citet{Smith2009} presented optical photometric and spectroscopic observations for the first 3 years post-discovery whilst \citet{Fox2009} presented near- and mid-infrared (IR) observations over the same period. Following an initial linear decline in its light curve, the luminosity of SN~2005ip began to plateau. Ongoing interaction with dense circumstellar material maintained this constant luminosity for several years and also caused significant X-ray emission allowing SN~2005ip to be detected by both {\em Chandra} and {\em Swift} in X-ray bands \citep{Immer2007,Katsuda2014}.  A strong near-IR excess was identified by \citet{Fox2010} caused by dust emission at 936\,d post-discovery. They concluded that there were likely two components to the observed IR flux excess, one arising from newly-formed hot dust associated with the supernova and another component caused by a thermal echo from pre-existing circumstellar dust located in a shell approximately 0.03~pc from the progenitor star.  Optical spectra at this time exhibited an unusual forest of high-ionization coronal lines and a blue pseudo-continuum that faded at later epochs \citep{Smith2009}. 

SN~2005ip is embedded in dense CSM and, as the forward shock impacted the surrounding material, a radiative reverse shock was propagated back through the ejecta creating a post-shock region of mixed ejecta shocked by the reverse shock and CSM shocked by the forward shock. The broad Balmer lines ($>15,000$~km~s$^{-1}$)  seen at early epochs emitted by the fast-moving ejecta therefore came to be dominated, after a few hundred days, by strong intermediate width lines  ($\sim2000$~km~s$^{-1}$) emitted by the post-shock gas. Narrow line profiles ($\sim 100$~km~s$^{-1}$) arising in the unshocked circumstellar material that was heated and ionised by the UV flash can often be seen superimposed upon intermediate and broad line profiles to produce complex line profile structures \citep{Smith2009,Stritzinger2012}.   

At later times, continued multiwavelength photometric observations in combination with visual spectra supported the assertion that two dust temperature components associated with pre-existing and newly-formed dust contributed to the observed near-IR flux \citep{Stritzinger2012}. \citet{Smith2009} suggested, based on observations in the first few years, that the progenitor system of SN~2005ip was a normal red supergiant (RSG) with a dense, clumpy wind comparable to VY Canis Majoris \citep{Smith2009a}. This `shellular' (i.e. concentric shells), roughly spherical geometry was also advocated by \citet{Stritzinger2012}. \citet{Smith2017} suggested that the SN blast wave was still interacting with a dense, clumpy CSM as late as early 2016, contrary to analysis of the X-ray flux by \citet{Katsuda2014} who claim that the forward shock has escaped the CSM shell. Most recently, \citet{Nielsen2018} used the light curve evolution of SN~2005ip to determine dust optical depths, concluding that dust formation had occurred in SN~2005ip starting as early as two months after discovery.

\subsection{Optical and near-IR X-shooter spectra at 4075 d}
We obtained observations of SN~2005ip in service mode with the X-shooter instrument mounted on the ESO VLT's UT2 Kueyen telescope \citep{Vernet2011}, on the nights of 
2016-12-27 and 2016-12-31 as part of a larger X-shooter survey of supernovae decades after outburst (PI M. J. Barlow, 097.D-0525(A)). The instrument was used in IFU staring mode. The total exposure time was 5729s in the UVB arm (3000--5000{\AA}), 5259s in the VIS arm (5000--10000{\AA}) and 6000s in the NIR arm (1.0--2.5\,$\mu$m), split into 5, 5 and 30 individual exposures respectively.

The data were reduced using EsoReflex \citep{Freudling2013}.  Basic 
calibrations were applied using dark, bias, arc and pinhole exposures 
taken within 24\,hr of the observations, and the spectra were 
flux-calibrated using observations of the spectrophotometric standard 
star LTT3218 taken on the same night.

EsoReflex does not carry out cosmic ray rejection for data taken in IFU 
staring mode, and so this was performed separately using a {\sc python} 
implementation of the Laplacian edge detection algorithm LACosmic \citep{vanDokkum2001}. From the fully reduced data cubes, we extracted the 
supernova spectrum from an aperture with a radius of 1'', after 
subtracting from all pixels within this radius a sky spectrum calculated 
as the median flux value of the pixels outside the aperture.

The NIR suffered some contamination by hot pixels which appear as very narrow lines in the reduced spectrum. These were removed from the intermediate-width He~{\sc i}~10830\,\AA\ line using a robust locally weighted smoothing (LOESS) routine to fit the line and identifying all lines of negligible width above a certain flux threshold.

Details of the wavelength range and resolution are given in Table \ref{tb_data} and the full spectrum is presented is Figure \ref{fig_X-shooter_spectrum}. We adopt the post-discovery date of 2016-12-31 (4075~d) for our combined X-shooter spectrum.

\subsection{Archival spectra}

\begin{figure*}
	\includegraphics[clip = true, trim = 20 0 20 0, width = \linewidth]{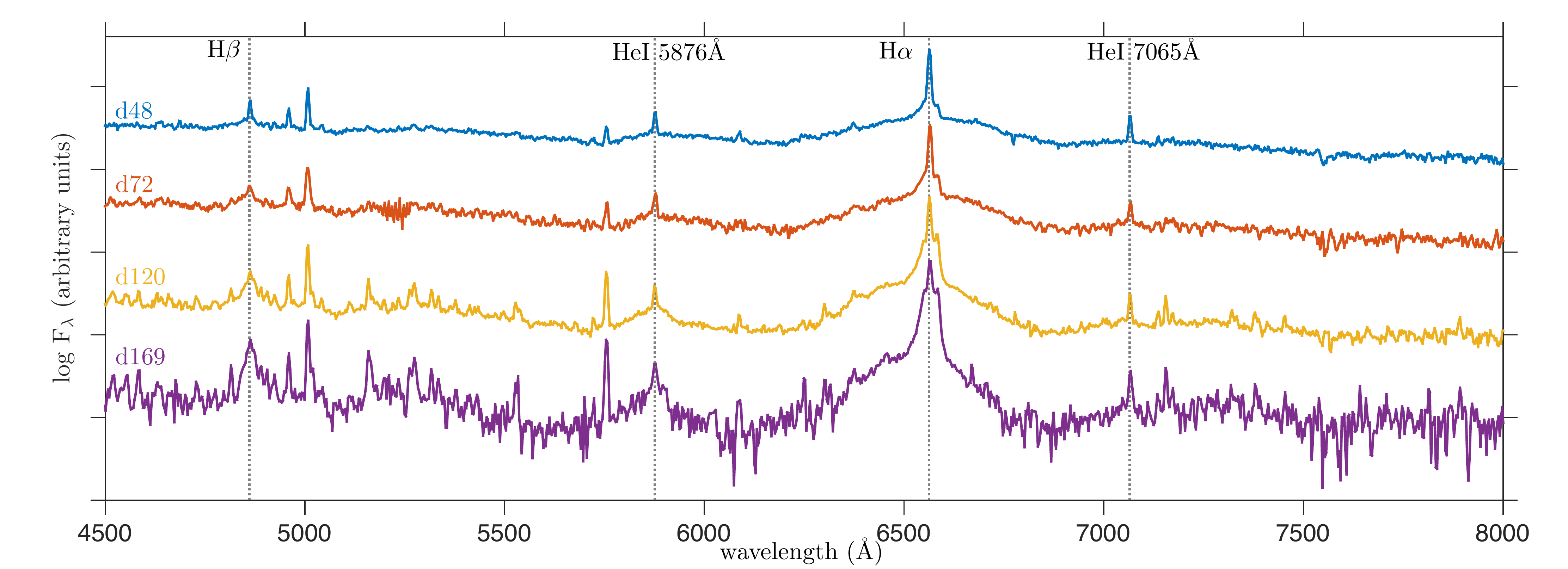}

	\includegraphics[clip = true, trim = 20 7 20 0, width = \linewidth]{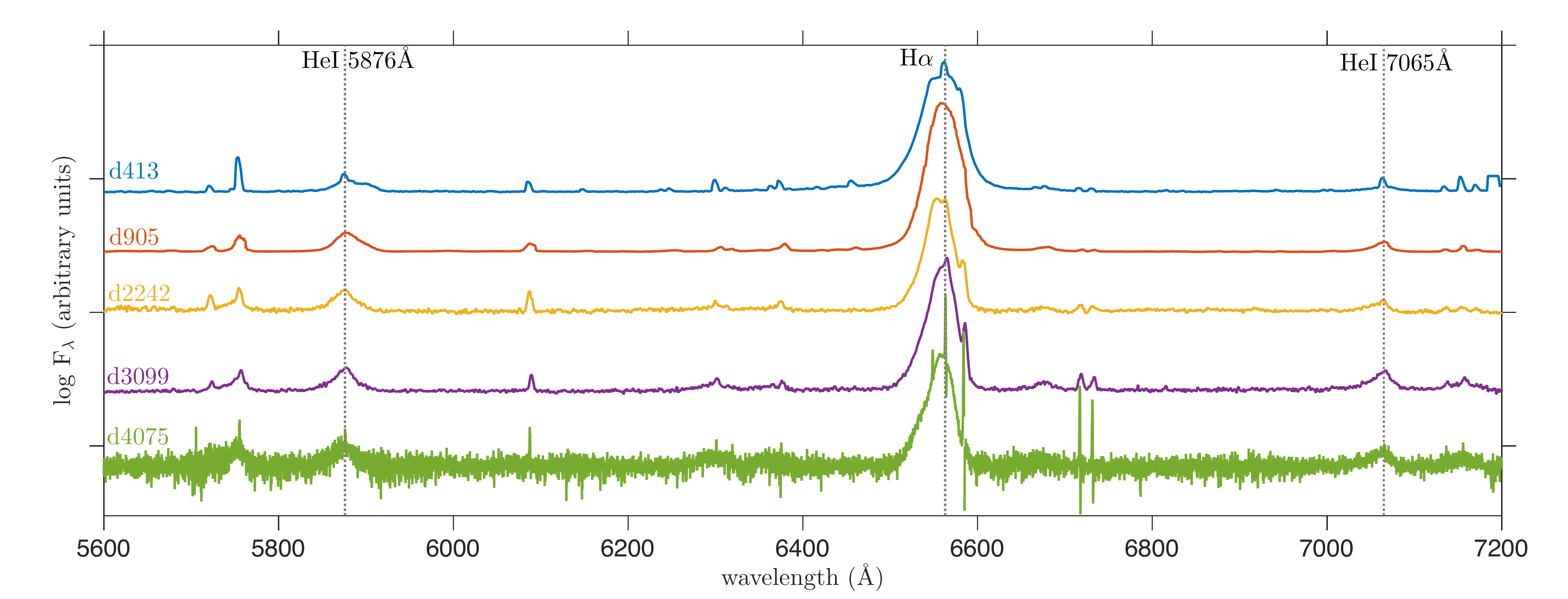}
	\caption{The evolution of the optical spectrum of SN~2005ip from 48~d to 4075~d. Details of the instruments and references for these spectra are presented in Table \ref{tb_data}. The vertical, dashed, black lines represent significant broad or intermediate width emission lines in the spectrum.}
	\label{fig_spectra}
\end{figure*}

To supplement our late-time spectrum of SN 2005ip we obtained optical spectra covering the evolution of the optical line profiles from shortly after outburst to the present day (Figure \ref{fig_spectra}). Spectra of the object taken during the first 200\,d were obtained from the Weizmann interactive supernova data repository (\citet{Wiserep}, http://wiserep.weizmann.ac.il). Optical spectra were acquired for 48\,d, 72\,d, 120\,d and 169\,d post-outburst with details and references given in Table \ref{tb_data}.

We also obtained archival spectra taken at 413\,d, 905\,d, 2242\,d and 3099\,d,  in addition to our X-shooter spectrum at 4075\,d post-discovery. Details and references are described in Table \ref{tb_data}. This gave a total of four early epoch spectra ($t<400$~d) and five later epoch spectra ($t>400$~d). These spectra were selected from a superset of publicly available spectra in order to provide a wide temporal coverage with reasonable sampling frequency. Where multiple spectra taken at similar epochs were available, spectra were  selected on grounds of spectral resolution, wavelength coverage and signal-to-noise ratio. Data obtained from the archives of the Keck observatory were reduced using the {\sc lpipe} pipeline for the reduction of LRIS spectra.

\renewcommand{\arraystretch}{1.1}
	\renewcommand{\tabcolsep}{0.15cm}
\begin{table*}
	\caption{Details of the spectral data for SN 2005ip.}
	\label{tb_data}
  	\begin{tabular*}{\linewidth}{L{2.2cm} C{1.6cm}C{1.6cm}C{1.6cm}C{1.6cm}C{1.6cm}C{1.6cm} l}
    	\hline
Date & Age & Telescope  & Inst & $\lambda$ Range & Res. & Res. Power & Reference \\
	& (d) & & &(\AA) & (FWHM\,\AA)& \\
	\hline
 23 Dec 2005 & 48 &  duPont & WFCCD & 3800-9235 & 8 & & \citet{Stritzinger2012} \\
 16 Jan 2006 & 72 &  NTT & EMMI & 4000-10200 & 8 & & \citet{Stritzinger2012} \\
 05 Mar 2006 & 120 &  duPont & WFCCD & 3800-9235 & 8 & & \citet{Stritzinger2012} \\
  23 Apr 2006 & 169 &  duPont & WFCCD & 3800-9235 & 8 & & \citet{Stritzinger2012} \\
    23 Dec 2006 & 413 &  Keck & DEIMOS & 4000-10500 &  & 2500 & \citet{Smith2009} \\
    28 Apr 2008 & 905 &  Keck & LRIS & 3200-10000 &  & 1500 & \citet{Smith2009} \\
    26 Dec 2011 & 2242 &  Keck & LRIS & 3200-10000 &  & 4500 & \\
    01 May 2014 & 3099 &  Keck & LRIS & 3200-10000 &  & 1500 & \citet{Smith2017} \\
    31 Dec 2016 & 4075 &  VLT & X-shooter & 3000-24800 &  & 7900-12600 & This work \\
    \hline
  \end{tabular*}
\end{table*}

\section{A Bayesian Approach to Monte Carlo Radiative Transfer}
\label{scn_model_setup}

We present here Monte Carlo radiative transfer models of optical and near-IR line profiles at  48, 72, 120, 169, 413, 905, 2242, 3099, and 4075\,d post-discovery using the method of \citet{Bevan2018}. We used the Monte Carlo line transfer code {\sc damocles} in combination with the {\sc python} Markov Chain Monte Carlo (MCMC) fitting routine {\em emcee} \citep{Goodman2010,emcee,damocles}. 

{\sc damocles} is a Monte Carlo line transfer code treating dust absorption and scattering effects for arbitrary geometrical distributions of dust and gas \citep{Bevan2016,damocles,Bevan2018}. Independent clumped or smooth distributions of dust and gas can be specified in addition to arbitrary velocity distributions. In all the models presented here, we adopted a geometry based on clumpy spherical shells of dust and gas motivated by the evidence that SN~2005ip had a roughly spherical, clumpy wind consistent with a RSG progenitor \citep{Smith2009,Stritzinger2012,Smith2017}. Further details of the model structures are given in Sections \ref{scn_model_early} and \ref{scn_model_late}.

{\em Emcee} is an affine invariant MCMC ensemble sampler that uses multiple walkers to sample the posterior probability distribution \citep{emcee}. It follows the algorithm described by \citet{Goodman2010} to determine the paths of the walkers and to efficiently and rigorously sample the posterior distribution, defined by Bayes' theorem as

\begin{equation}
\label{bayes_thm}
P(\boldsymbol \theta\,|\,\boldsymbol D) \propto P(\boldsymbol \theta ) \, P(\boldsymbol D\,|\,\boldsymbol \theta)
\end{equation}

\noindent where $\boldsymbol D$ is the data, $ \boldsymbol \theta$ is the set of parameters of the model, $P(\boldsymbol \theta )$ is our prior understanding of the probability of the parameters (the prior), and $P(\boldsymbol D\,|\,\boldsymbol \theta)$ is the probability of obtaining the data for a given set of  parameters (the likelihood).

The final posterior probability distribution is presented as a series of 2D contour plots for each pair of parameters marginalised over the other parameters. In addition, a 1D marginalised posterior probability distribution for each parameter is presented. Many of the 1D marginalised probability distributions deviate strongly from a Gaussian distribution.  We therefore adopted the modal parameter value with uncertainties defined by the region of highest probability density as the metric most representative of the distribution.  We adopted a credibility region of 68\% for all parameters.

Our formulation of the likelihood function follows \citet{Bevan2018} where the likelihood function is proportional to $\exp({-\chi_{\rm red}^2/2})$ and   $\chi_{\rm red}^2$ is defined as 

\begin{equation}
\label{eqn_chi2}
\chi^2_{\rm red} = \frac{\chi^2}{\nu}, \quad \quad \chi^2 = \sum^n_{\rm i=1} \frac{(f_{\rm mod,i}-f_{\rm obs,i})^2}{\sigma_{\rm i}^2}
\end{equation}

\noindent where $\nu$ is the number of degrees of freedom, $f_{\rm obs,i}$ is the observed flux in frequency bin $i$, $f_{\rm mod,i}$ is the scaled, modelled flux in bin $i$ and $\sigma_{\rm i}$ is the combined Monte Carlo and observational uncertainty in bin $i$ \citep{Bevan2018}.

For all models, a 7-dimensional variable parameter space was investigated using 500 walkers. The observational uncertainty for each spectrum was estimated by taking several regions of flat continuum that were not contaminated by any form of line emission and calculating the variance of the flux in these regions. All models were run to convergence as determined by the evolution of the auto-correlation and were run for a minimum of 6000 steps. In all cases, it was required that the acceptance fraction be $>0.15$ but in the majority of cases the acceptance fraction was in the range $0.25 - 0.3$.

Uniform priors were adopted for all parameters with the exception of the dust mass and dust grain radius. Unlike the other parameters (see Section \ref{scn_all_model_results} for details), the prior range of these two parameters spanned several orders of magnitude such that a log-uniform prior was a more appropriate choice. For all line profiles, we first modelled the profiles manually, tuning the parameters by eye in order to establish reasonable prior bounds and starting distributions for the walkers.  Even so, prior ranges were kept as wide as was computationally feasible. Runs were performed on a local cluster using multiple cores (between 5 and 20) and generally took several days to converge. A summary of the variable parameters for each epoch and the associated priors can be found in Table \ref{tb_priors}.

\renewcommand{\arraystretch}{1.1}
	\renewcommand{\tabcolsep}{0.15cm}
\begin{table*}
	\caption{Details of the adopted prior ranges for the early epochs ({\em top}) and later epochs ({\em bottom}). $v_{\rm max}$ and $v_{\rm min}$ are respectively the maximum and minimum radial velocity of the ejecta, $\beta_{\rm clump}$ is the index of the radial power-law dust clump number distribution, $\beta_{\rm gas}$ is the smooth gas density radial power-law distribution (on which the emissivity distribution depends), $\alpha_{\rm vel}$ is the index of the velocity power-law probability distribution, $f_V$ is the fraction of the ejecta volume occupied by dust clumps, $\log M_{\rm d}$ is the $\log$ of the dust mass. and  $\log a$ is the $\log$ of the single grain radius.}
	\label{tb_priors}
  	\begin{tabular*}{\linewidth}{C{0.95cm} C{1.7cm}C{1.7cm}C{1.4cm}C{1.4cm}C{1.4cm}C{1.4cm}C{1.4cm}C{1.4cm} L{2.5cm}}
    	\hline
 Epoch  & $v_{\rm max}$  & $v_{\rm min}$ & $\beta_{\rm clump}$  & $\beta_{\rm gas}$ & $f_V$ & $\log M_{\rm d}$  & $\log a$ &Model geometry&Broad line modelled\\
	  (d) & (10$^3$~km~s$^{-1}$)  & (10$^3$~km~s$^{-1}$) &   &  &  & ($\log$ M$_{\odot}$)  & ($\log$ \si{\um}) &\\
     \hline
  48&[15.0,19.0]&[1.0,6.0]&[0.0,4.0]&[0.0,2.5]&[0.05,0.8]&[-9.0,-5.0]&[-2.0,0.7]&Fig. \ref{fig_graphic_early}&Broad H$\alpha$\\
72&[12.0,18.0]&[1.0,6.0]&[0.0,4.0]&[0.0,2.5]&[0.05,0.8]&[-9.0,-4.0]&[-2.0,0.7]&Fig. \ref{fig_graphic_early}&Broad H$\alpha$\\
120&[9.0,15.0]&[1.0,6.0]&[0.0,4.0]&[0.0,2.5]&[0.05,0.8]&[-7.0,-3.0]&[-2.0,0.7]&Fig. \ref{fig_graphic_early}&Broad H$\alpha$\\
169&[10.0,16.0]&[1.0,4.0]&[0.0,4.0]&[0.0,2.5]&[0.05,0.8]&[-7.0,-3.0]&[-2.0,0.7]&Fig. \ref{fig_graphic_early}&Broad H$\alpha$\\
\hline
\\
\hline
 Epoch  & $v_{\rm max}$  & $v_{\rm min}$ &  $\alpha_{\rm vel}$ & $\beta_{\rm gas}$  & $f_V$ & $\log M_{\rm d}$  & $\log a$ &Model geometry&Intermediate line(s) modelled\\
	  (d) & (10$^3$~km~s$^{-1}$)  & (10$^3$~km~s$^{-1}$) &   &  &  & ($\log$ M$_{\odot}$)  & ($\log$ \si{\um}) &\\
	\hline
    413&[1.0,4.0]&[0.05,0.8]&[-3.0,3.0]&[0.0,10.0]&[0.05,0.8]&[-6.0,-0.5]&[-2.0,0.7]&Fig. \ref{fig_graphic_late_ej_dust}&He{\sc i}~7065\AA \\
905&[1.0,4.0]&[0.05,0.8]&[-3.0,3.0]&[0.0,10.0]&[0.05,0.8]&[-6.0,-0.5]&[-2.0,0.7]&Fig. \ref{fig_graphic_late_ej_dust}&He{\sc i}~7065\AA \\
2242&[1.0,4.0]&[0.05,0.9]&[-3.0,3.0]&[0.0,10.0]&[0.05,0.8]&[-6.0,0.2]&[-2.0,0.7]&Fig. \ref{fig_graphic_late_ej_dust}&H$\alpha$, H$\beta$, He{\sc i}~7065\AA\\
3099&[1.0,3.0]&[0.05,0.9]&[-1.5,3.0]&[0.0,10.0]&[0.05,0.8]&[-6.0,0.2]&[-2.0,0.7]&Fig. \ref{fig_graphic_late_ej_dust} &H$\alpha$, H$\beta$, He{\sc i}~7065\AA\\
4075&[1.0,3.0]&[0.05,0.9]&[-1.5,3.0]&[0.0,10.0]&[0.05,0.8]&[-6.0,0.2]&[-2.0,0.7]&Fig. \ref{fig_graphic_late_ej_dust}&H$\alpha$, H$\beta$, He{\sc i}~10830\AA\\
     \hline
  \end{tabular*}
\end{table*}

\section{Line profile models}
\label{scn_all_model_results}

We  modelled the early-time evolution of the broad ($\sim$15000\,km~s$^{-1}$) H$\alpha$ line from an almost symmetric line profile at 48~d to a distinctly asymmetric, blueshifted line at 169~d. The broad H$\alpha$ line arises in the fast-moving ejecta but rapidly becomes dominated by the intermediate width  H$\alpha$ line originating in the post-shock region. It is therefore not possible to trace the broad line component after this time (see Figure \ref{fig_spectra}).

At later epochs we presumed that the intermediate width lines ($\sim$2000 km~s$^{-1}$) were emitted from excited gas in the post-shock region between the forward and reverse shocks (Figure \ref{fig_model_graphic}). Intermediate width lines could also arise as the photosphere recedes and reveals slower moving material but the expected timescales for this mechanism are considerably shorter \citep[$\sim$50d; e.g.][]{Woosley1988} than the epoch of the first appearance of the intermediate width He\,{\sc i} lines ($\sim$413\,d).

\begin{figure}
	\centering
	\includegraphics[clip = true, trim = 20 0 30 20, width = \linewidth]{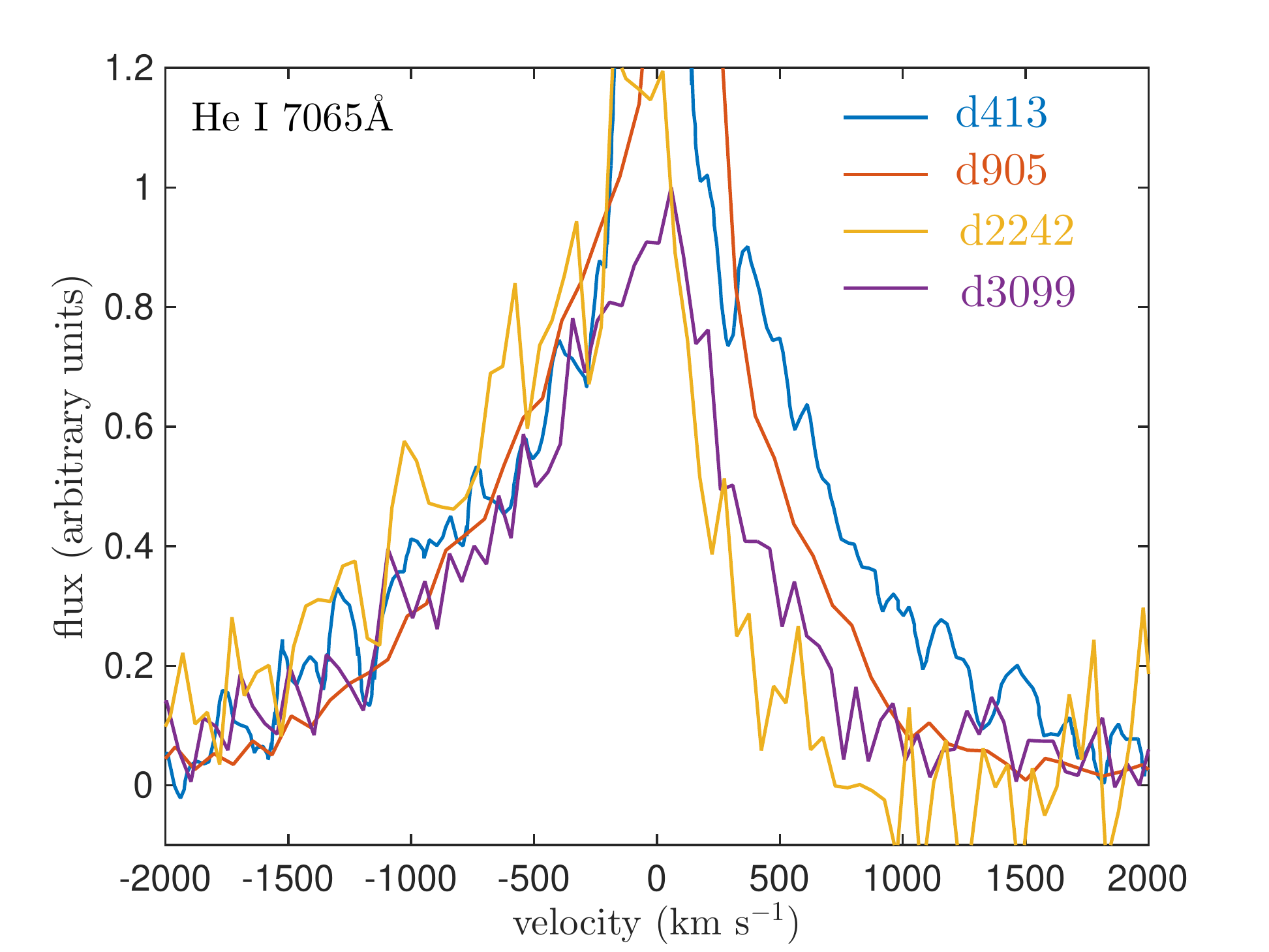}
	\caption{The evolution of the He~{\sc i}\,7065\,\AA\ line at later epochs showing the increasing attenuation of the red side of the profile. All profiles are continuum-subtracted and scaled to the peak flux of the intermediate width component.}
	\label{fig_7065}
\end{figure}

At 413~d and 905~d, the strongest line in the spectrum was H$\alpha$. However, the confused broad and intermediate width contributions from  the fast-moving ejecta and the slower post-shock region result in a complicated intrinsic emissivity distribution. We instead focused solely on the He~{\sc i}~7065~\AA\ line at these epochs which does not exhibit a broad ejecta component. At days 2242, 3099 and 4075~d, however, the broad component of the Balmer lines has faded significantly leaving only the intermediate width component and allowing us to model the H$\alpha$, H$\beta$, and He~{\sc i}~7065~\AA\ or He~{\sc i}~10830~\AA\ lines simultaneously. The evolution of the He~{\sc i}~7065~\AA\ line at later epochs is presented in Figure \ref{fig_7065}.

We have not modelled the relatively strong He~{\sc i}~5876~\AA\ at any epoch due to contamination by Na~{\sc i}~D on the red side of the profile that is too blended to distinguish the He~{\sc i}~5876~\AA\ profile alone \citep{Smith2009}.

In all models presented here, we use amorphous carbon grains and the optical constants of the BE sample of \citet{Zubko1996}. This is motivated by the absence of any silicate features in the mid-IR spectrum \citep{Fox2010}. From previous models, silicate dust could increase the inferred dust mass by as much as an order of magnitude. We discuss the possible impact of silicate dust further in Section \ref{scn_discuss_87A_comparison}. 

For the purposes of this paper, we have only considered the possibility of dust formation in the ejecta of SN~2005ip in order to test whether ejecta dust alone can account for the observed line profiles. We adopted a shell-based geometry with a shellular line emitting region defined by inner and outer radii $R_{\rm in}$ and $R_{\rm out}$. Dust was located in clumps distributed throughout an ejecta shell. In the case of the early-time, broad line models where emission originates in the ejecta, these shells are coincident. In the case of later time, intermediate width lines where the emission arises in a post-shock region, the line emitting region is exterior to the dusty shell in the models.  Further details of the geometries adopted for the broad and intermediate width line models are given in Sections \ref{scn_model_early} and \ref{scn_model_late} respectively.

Details of which lines were modelled for each epoch, along with the adopted priors are summarised in Table \ref{tb_priors}. 

\subsection{The broad H\texorpdfstring{\textalpha{}}{alpha} line \texorpdfstring{at t \textless{} 400\,d}{before 400 d}}
\label{scn_model_early}

The broad H$\alpha$ line is present in the spectrum of SN~2005ip from very early times (Figure \ref{fig_spectra}). It exhibits velocities $>15,000$~km~s$^{-1}$ and originates in the fast-moving ejecta. Initially, the profile is symmetric (48~d). However, over the next $\sim$100~d, the red side of the profile becomes increasingly attenuated (see \citet{Smith2009} Fig. 6).  This evolving asymmetry can be simply explained by dust formation in the ejecta of SN~2005ip during its early phases. Similar asymmetries attributed to dust formation have been observed at these early times for the similar Type IIn SN 2010jl \citep[][Krafton in prep.]{Gall2014} and have also been observed at much later times ($\gtrsim 500$\,d) in non-interacting CCSNe such as SN~1987A \citep{Lucy1989,Milisavljevic2012,Bevan2017}. 
The red-blue asymmetry appears  in the broad lines of SN~2005ip and persists in multiple intermediate width lines at later epochs which further supports dust formation as the likely origin of the asymmetry in the broad H$\alpha$ line at early epochs.

The attenuation of line radiation by dust  will result in a red-blue asymmetry of the line only when the dust is either internal to or co-located with the source of the line emission.  Dust that is external to the radiation source (such as interstellar or pre-existing circumstellar dust) will attenuate radiation uniformly across the width of the line. No asymmetry would be observed in this case.  At these early epochs therefore, when lines arise in the fast-moving ejecta, dust must be forming in the ejecta. 

\subsubsection{Model details}
In our models of the broad H$\alpha$ line (at epochs $<400$~d), we treated emission emanating from a shell defined by an inner radius $R_{\rm in}$ and an outer radius $R_{\rm out}$. $R_{\rm in}$ and $R_{\rm out}$ were not fixed but were varied via the parameters $v_{\rm min}$ and $v_{\rm max}$ (the minimum and maximum radial velocities of the ejecta respectively), appropriate for a shell in homologous expansion. $v_{\rm max}$ and $v_{\rm min}$ were sampled from a uniform prior with the range for each epoch based on a visual inspection of the line profiles (see Table \ref{tb_priors}).

Dust was assumed to form in the ejecta and was therefore colocated with the gas, distributed between $R_{\rm in}$ and $R_{\rm out}$. Both gas and dust were assumed to follow a velocity law $v \propto r$ appropriate for freely-expanding ejecta. The emissivity distribution was proportional to the square of the gas density distribution which was governed by the free parameter $\beta_{\rm gas}$ such that the smooth gas density distribution was $\propto r^{-\beta_{\rm gas}}$. 

\begin{figure*}	
	\centering	
	\subfloat[Broad line emitting model at early times]{\includegraphics[clip = true, trim = 170 190 190 110, width = 0.3\linewidth]{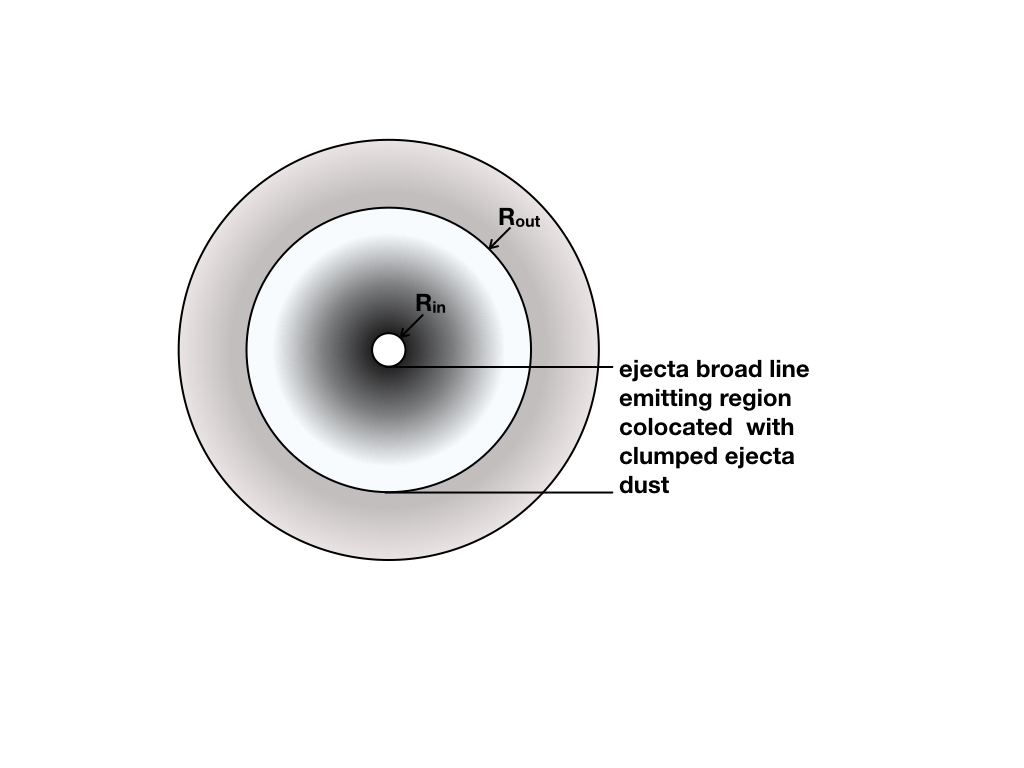}\label{fig_graphic_early}}\hspace{0.25cm}
	\subfloat[Intermediate width line emitting model at later times with clumped ejecta dust]{\includegraphics[clip = true, trim = 170 190 190 110, width = 0.3\linewidth]{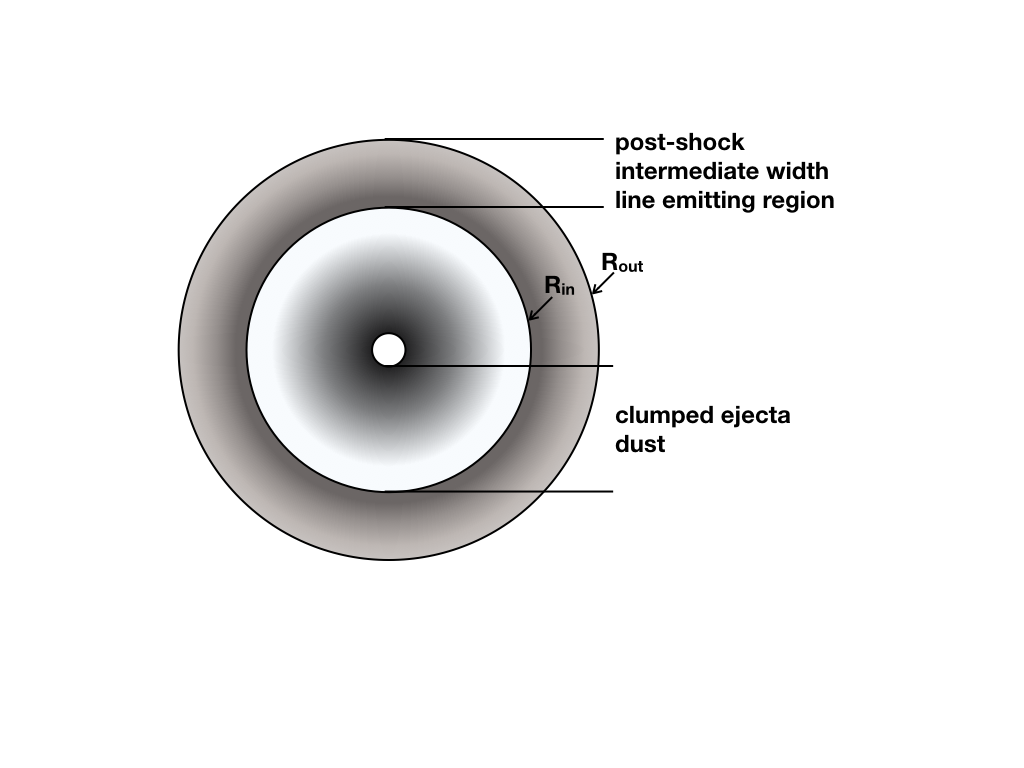}\label{fig_graphic_late_ej_dust}}
    \hspace{0.25cm}
    \subfloat[Intermediate width line emitting model at later times with clumped dust in the post-shock region]{\includegraphics[clip = true, trim = 170 190 190 110, width = 0.3\linewidth]{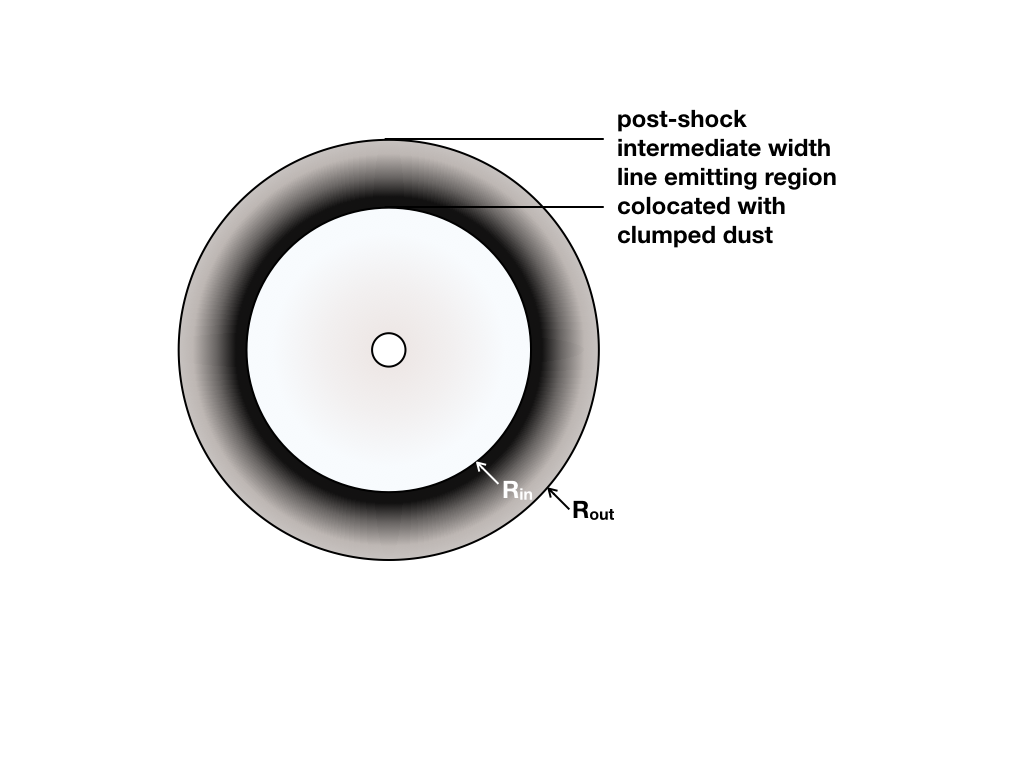}\label{fig_graphic_late_ps_dust}}
	\caption{ Possible dust formation locations in IIn SNe. (a) and (b) represent the model geometries used at early and later times respectively. Dust is assumed to form in the ejecta in these models. $R_{\rm in}$ and $R_{\rm out}$ represent the bounds of the line emitting shell (the ejecta in scenario (a) and the post-shock region in scenario (b)). The post-shock dust formation scenario (c) is presented for comparison (see Section \ref{scn_discussion} for a more detailed discussion).}
	\label{fig_model_graphic}
\end{figure*}

The dust distribution was independent of the gas distribution. We located 100\% of the dust in clumps of size $R_{\rm out}/25$ consistent with the expected scale of the Rayleigh-Taylor instabilities that are thought to give rise to these clumps \citep{Arnett1989}. 
The clumps were stochastically distributed between $R_{\rm in}$ and $R_{\rm out}$ according to a probability law $\propto r^{-\beta_{\rm clump}}$ where $\beta_{\rm clump}$ was a variable parameter. The total number of clumps was determined by the clump volume filling factor $f_{\rm V}$. The dust mass and single dust grain radius were also varied. Figure \ref{fig_graphic_early} illustrates the ejecta shell geometry adopted for these models and Table \ref{tb_priors} lists all of the variable parameters with their associated prior ranges.

\begin{figure}
\centering
\subfloat{\includegraphics[clip = true, trim = 0 8 0 5, width = 0.8\linewidth]{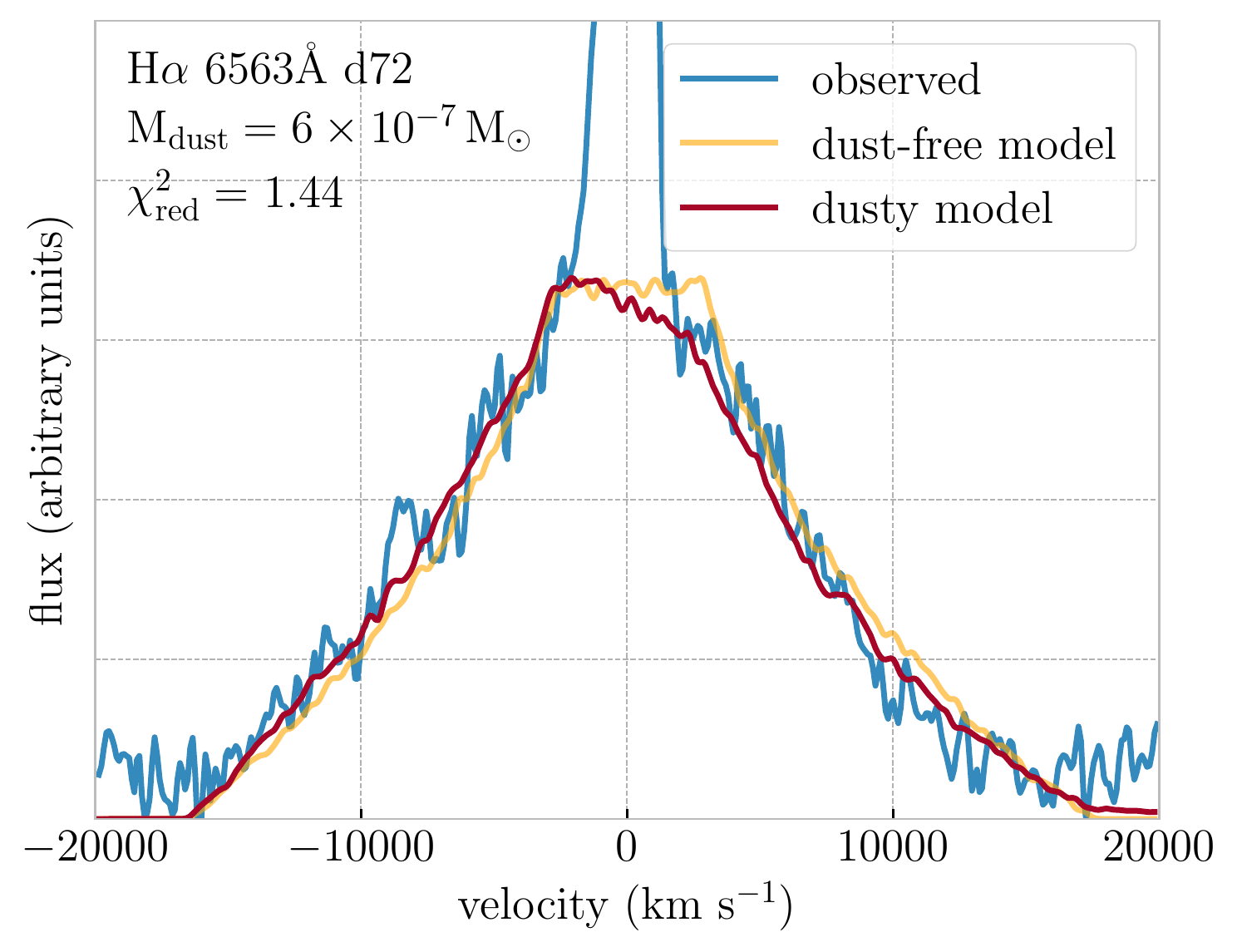}}

\subfloat{\includegraphics[clip = true, trim = 0 8 0 5, width = 0.8\linewidth]{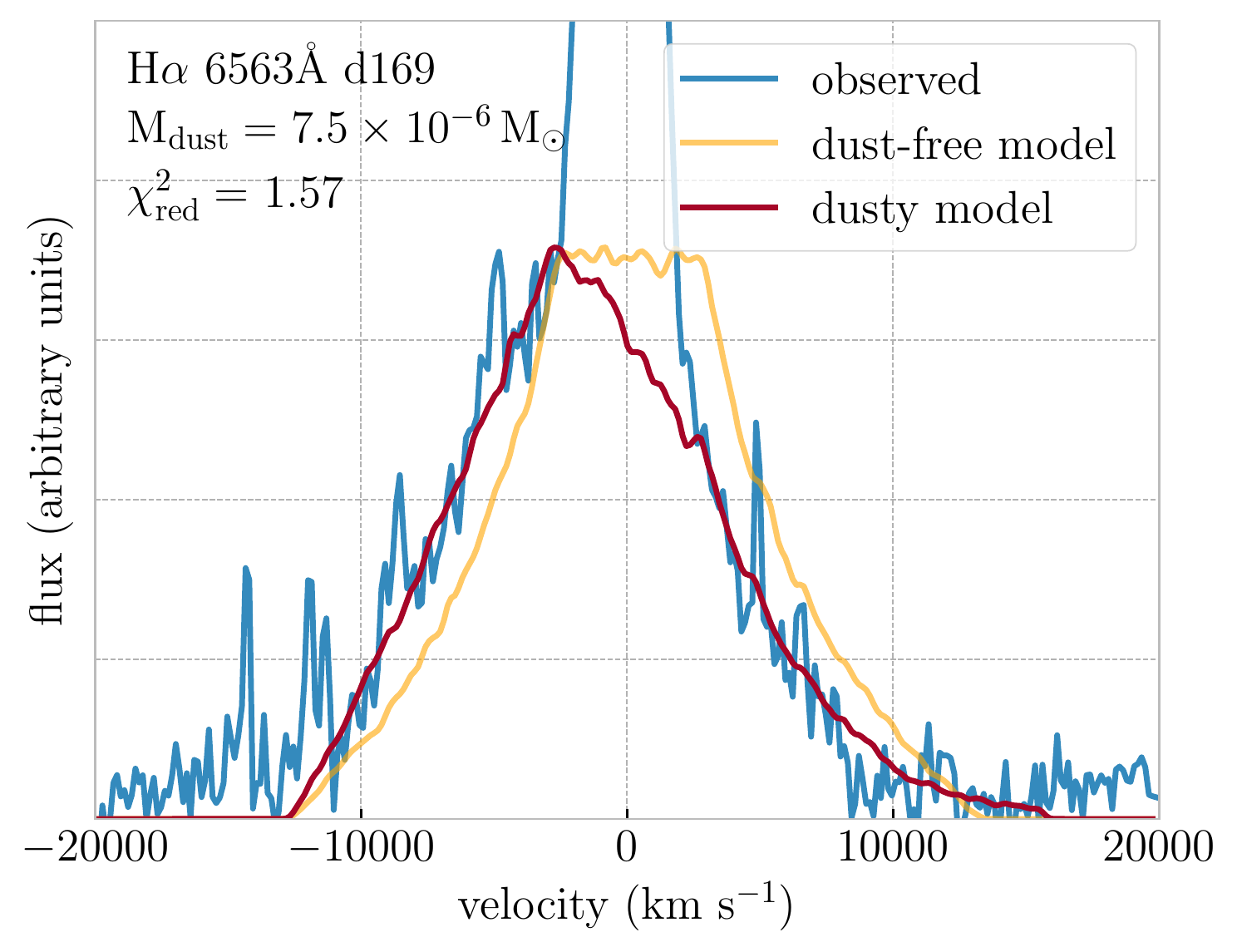}}
\caption{Examples of the best-fitting models for two of the early epochs. The observed spectrum is show in blue, the intrinsic dust-free profile (scaled to the peak flux of the best-fitting model) is shown in yellow and the best-fitting model profile is shown in red. Dust masses were determined using amorphous carbon grains of radius 0.08{\si{\um}}. Intermediate width emission was neglected for the purposes of fitting the broad line.}
\label{fig_early_best_fits}
\end{figure}

The adjacent continuum level was subtracted for each line and the portion of the profile contaminated by the intermediate and narrow H$\alpha$ emission was excluded from the calculation of $\chi^2_{\rm red}$. 
Before commencing a full parameter space investigation using the automated MCMC algorithm, we first identified the best-fitting set of parameters using a manual approach in a simplified scenario (dust was distributed smoothly and coupled to the gas density distribution). This was to judge the appropriateness of our dusty model and also to understand better the prior range that should be adopted for the Bayesian models.  We present examples of our best-fitting model line profiles for each epoch to illustrate the quality of the fits obtained (Figure \ref{fig_early_best_fits}).

\subsubsection{Ejecta dust formation at early times}
\label{early_model_results}
\begin{figure*}
	\includegraphics[width = 1\linewidth]{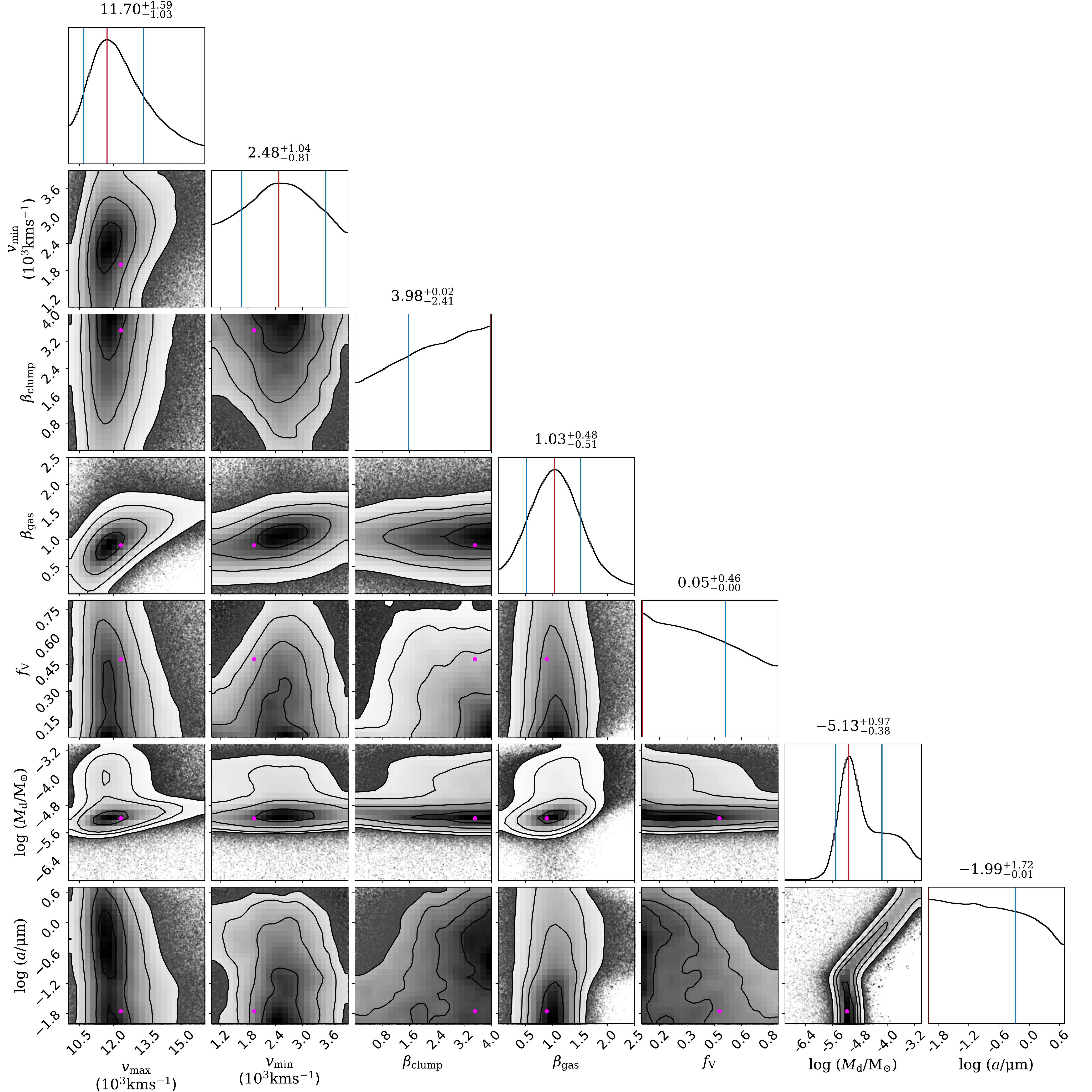}
	\caption{The full 7D posterior distribution for the MCMC model of the broad H$\alpha$ line of SN~2005ip at 169~d.  The vertical red line indicates the mode whilst the blue vertical lines indicate the bounds of the 68\% credibility region. The pink circle indicates the best-fitting parameter set from the MCMC run.}
	\label{fig_169_full_posterior}
\end{figure*}

The full 7D posteriors for all epochs are presented in the online material. We include the full posterior for the 169~d model here for illustration purposes (Figure \ref{fig_169_full_posterior}). The best-fitting set of parameters from the MCMC run is marked on the contour plots for comparison. The modal parameter value along with the 68\% credibility region for each 1D marginalised distribution for all parameters and epochs is presented in full in the Appendix in Table \ref{tb_results_full} with the results for dust mass and grain radius summarised in Table \ref{tb_dust_masses_early}.

\renewcommand{\arraystretch}{1.6}
	\renewcommand{\tabcolsep}{0.32cm}
\begin{table}
	\caption{Dust mass and grain radius estimates for the amorphous carbon models of the broad H$\alpha$ line profile of SN~2005ip at early times ($t<400$~d). The modal value from the marginalised 1D dust mass and grain radius distribution is given in addition to the 68\% confidence interval. A $^*$ indicates that the lower limit is equal to the lower bound of the prior range. }
	\label{tb_dust_masses_early}
    \centering
  	\begin{tabular*}{\linewidth}{C{0.4\linewidth} C{0.4\linewidth} }
	\hline
 Epoch  & $\log M_{\rm d}$   \\
(d)  & ($\log$ M$_{\odot}$)   \\
\hline
48 &$ -7.80 \substack{ +0.87 \\ -1.20 ^*}$ \\ 
72 &$ -6.54 \substack{ +0.46 \\ -2.10 ^*}$\\ 
120 &$ -5.51 \substack{ +1.01 \\ -0.39 }$\\ 
169 &$ -5.13 \substack{ +0.97 \\ -0.38 }$\\ 
\hline
  \end{tabular*}
\end{table}

We are primarily interested in understanding the evolution of the dust mass as dust grains form in the ejecta of SN~2005ip.  This is fundamentally dependent on the dust grain radius.  At 48~d and 72~d, the 1D marginalised posterior distributions indicate an apparent preference for larger grains. However, this trend is a product of the  adoption of a flat, 'uninformative' prior for the grain radius and a wide prior dust mass range. The marginalised 1D posteriors have, by definition, been integrated over all of the other parameters. Given that a wide range of dust masses have been explored, for any parameter set with a high dust mass, a larger (and less absorbing) grain size will generate a better fit to the nearly symmetric line profiles at 48\,d and 72\,d. Over the whole parameter space, there are more ‘good fitting’ models which have large grain radii than there are ‘good fitting’ models which have smaller radii resulting in a higher density of points in the larger grain radius region of the parameter space and an artificial preference (in the 1D posteriors) for larger grains. 

 We should rightly be skeptical of such a large grain size at such an early epoch, and ideally such knowledge should be incorporated into the prior distribution. However, quantifying this knowledge in the prior is not straightforward and could introduce other biases. We therefore adopted a log-uniform prior for the grain radius and highlight here the importance of considering the 2D posteriors. The 2D posterior clearly indicates that larger grain sizes are only preferred for larger dust masses.

\begin{figure}
\centering
\subfloat{\includegraphics[clip = true, trim = 0 0 0 0, width = 0.4\linewidth]{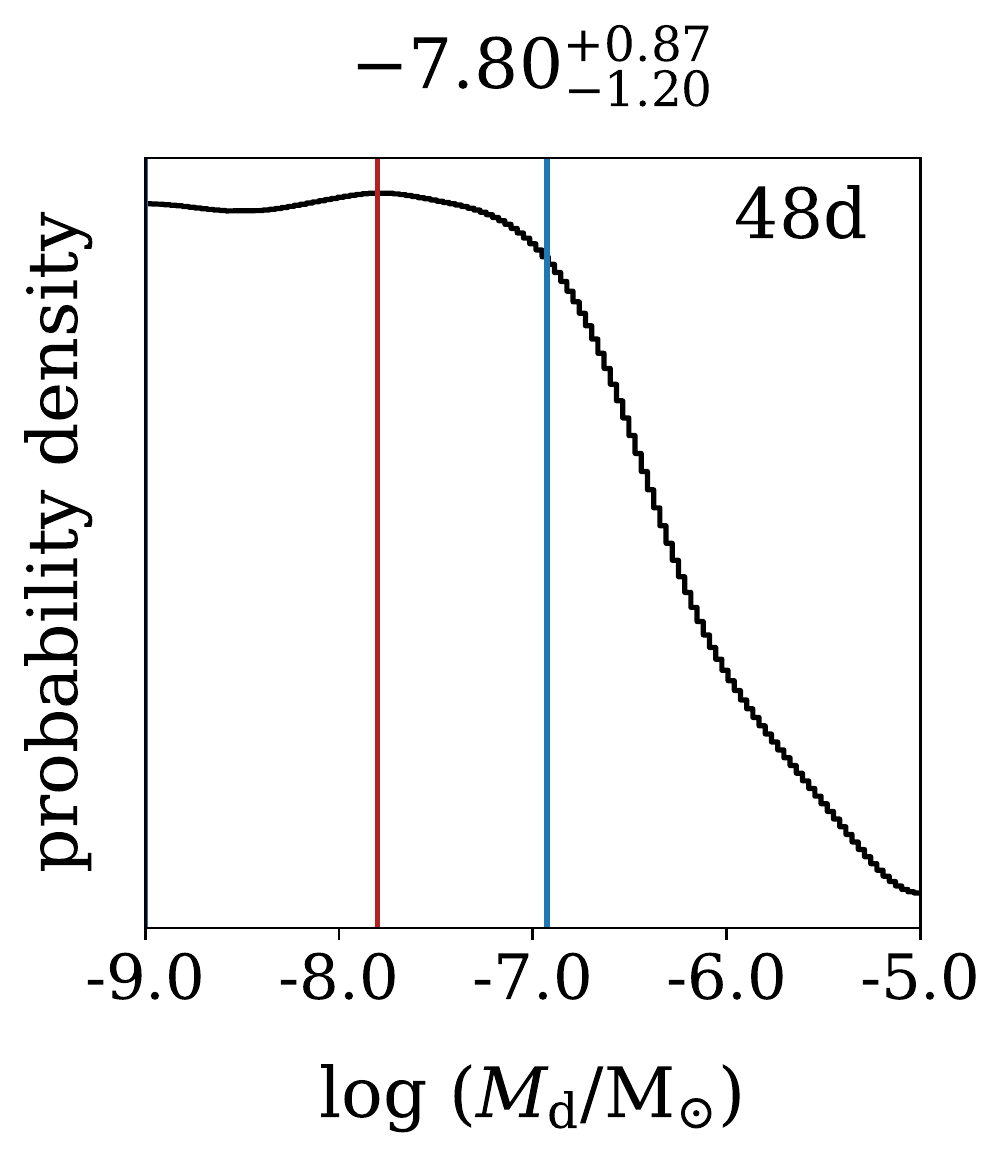}} \hspace{0.2cm}
\subfloat{\includegraphics[clip = true, trim = 27 0 0 0, width = 0.36\linewidth]{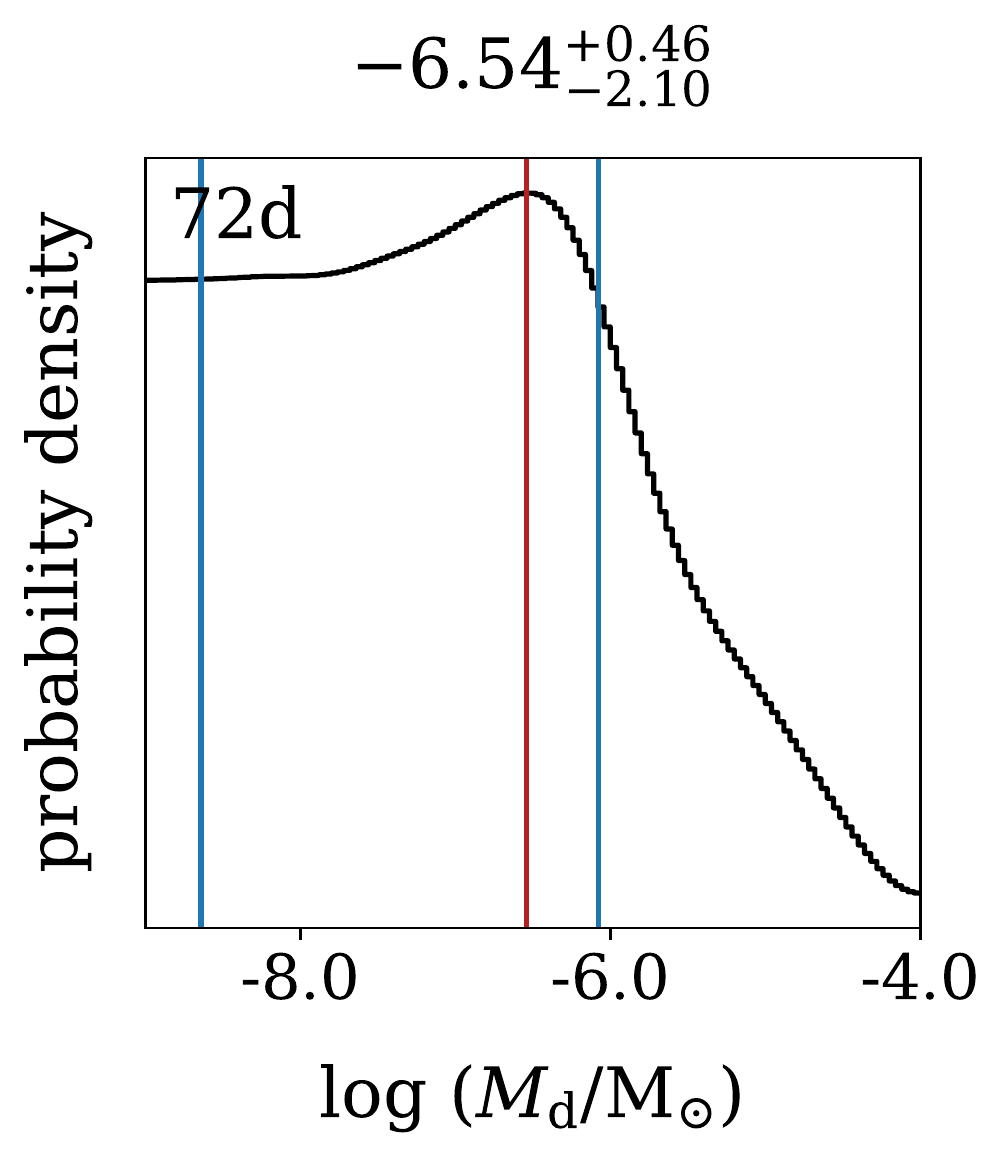}}

\subfloat{\includegraphics[clip = true, trim = 0 0 0 0, width = 0.4\linewidth]{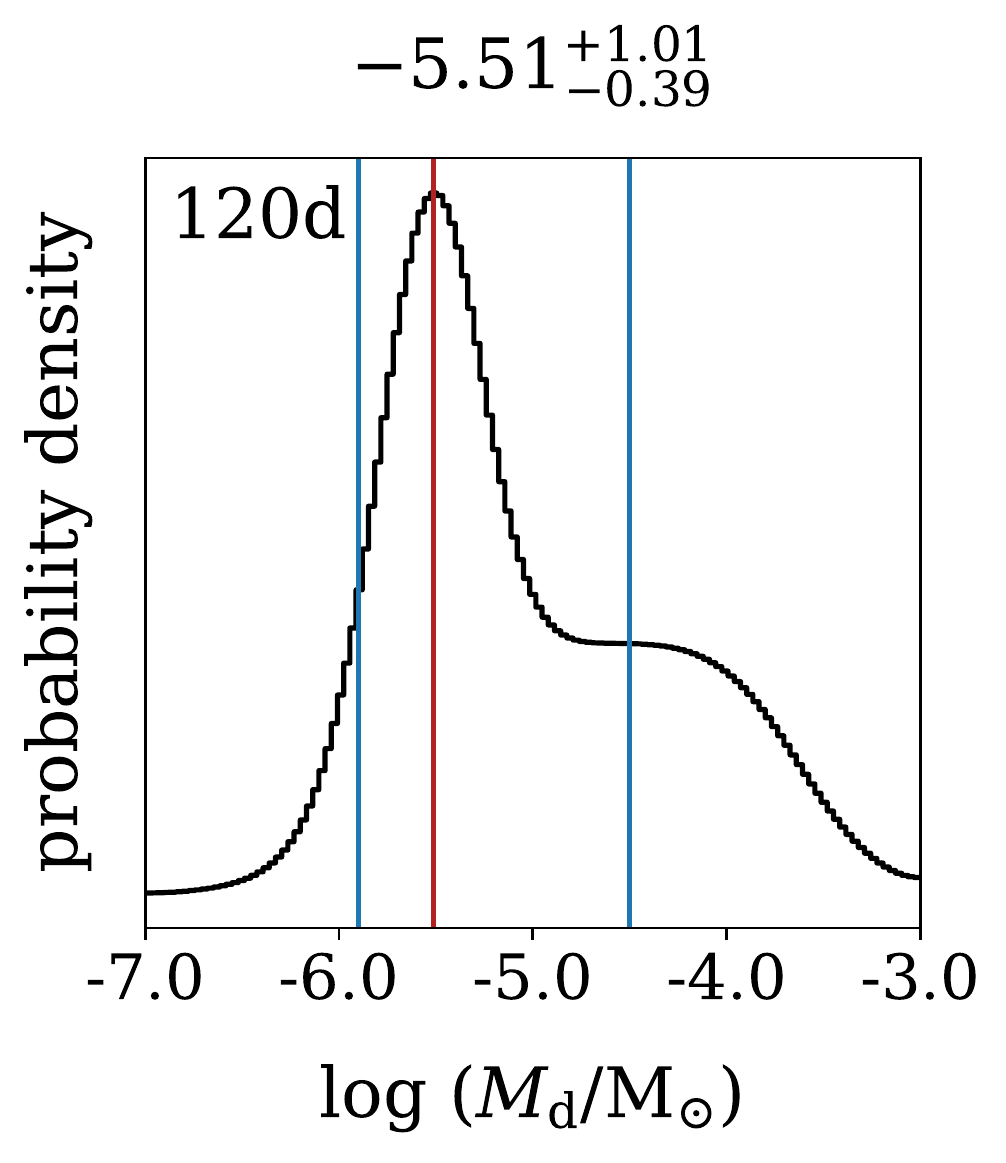}} \hspace{0.2cm}
\subfloat{\includegraphics[clip = true, trim = 27 0 0 0, width = 0.36\linewidth]{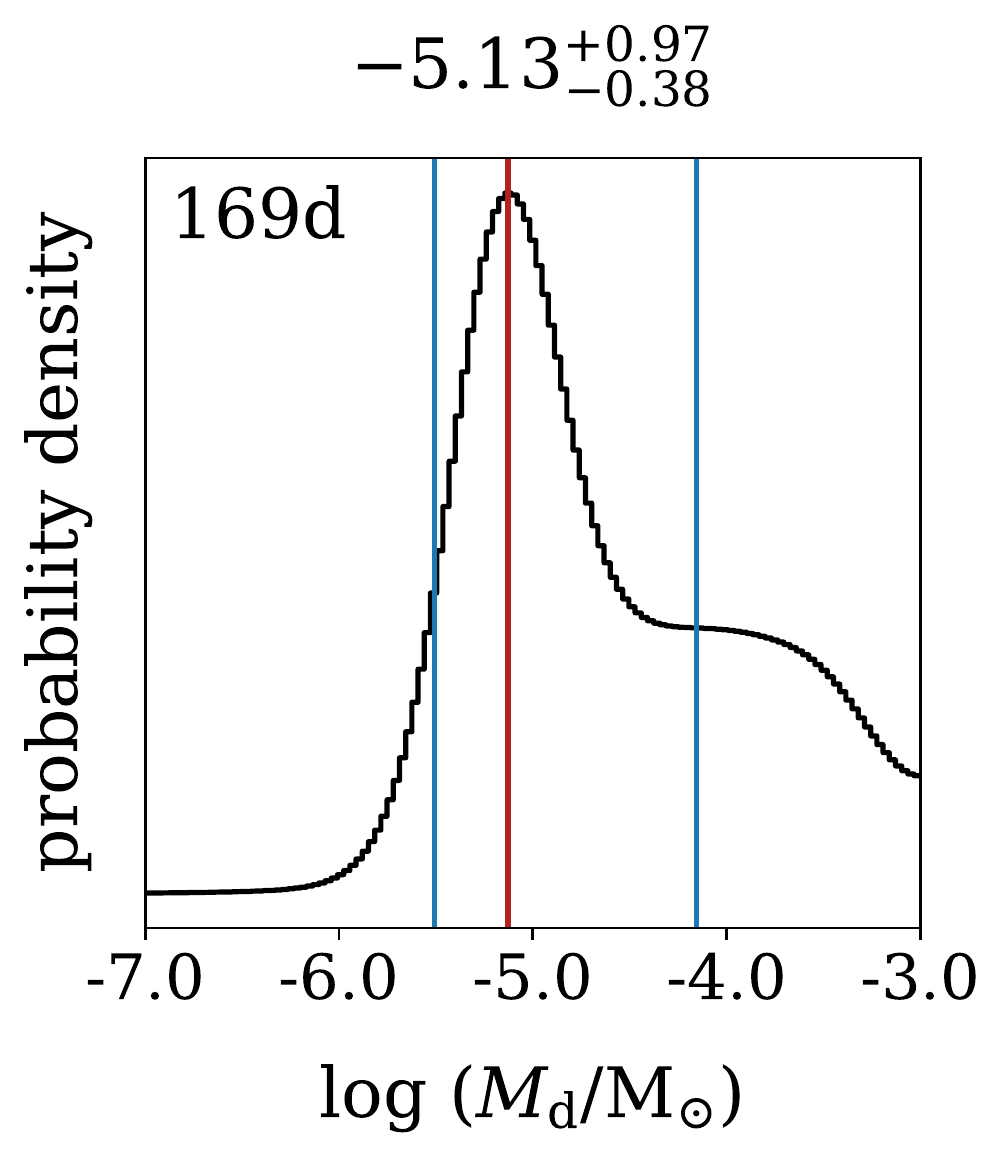}}
\caption{Dust mass 1D marginalised posterior probability distribution for the early-time models. Vertical blue lines indicate the upper and lower bounds of the shortest region containing 68\% of the probability density (i.e. the region of highest probability density). Red vertical lines indicate the modal dust mass.}
\label{fig_early_1D_dustmass}
\end{figure}

However, by marginalising over the grain radius (and the other parameters), reasonably strong constraints can be placed on the ejecta dust mass at early times. We present the 1D dust mass marginalised posterior probability distributions in Figure \ref{fig_early_1D_dustmass}.

The initially symmetric H$\alpha$ line profile at 48~d allows an upper limit of $10^{-7}$~M$_{\odot}$ to be placed on the dust mass (with more dust resulting in a red-blue asymmetry that is not observed). At both 48~d and 72~d, the dust mass probability distribution can be seen to fall off sharply, and the peak in the distribution for 72~d indicates an upper limit of $\sim 10^{-6}$M$_{\odot}$. An increasingly pronounced peak in the distribution appears over time allowing upper and lower limits to be placed on the ejecta dust mass, reaching $\log M_{\rm d} = -5.13 \substack{ +0.97 \\ -0.38 }$ by 169~d. We summarise this early-time dust mass evolution in Figure \ref{fig_dust_mass_evol}, along with the later time evolution derived from models of the intermediate width lines.

\subsection{Intermediate-width line models \texorpdfstring{at t \textgreater{} 400\,d}{after 400 d}}
\label{scn_model_late}
As the broad H$\alpha$ profile steadily becomes dominated by the intermediate width component, so the intermediate width He\,{\sc i} lines begin to appear. After a few hundred days, intermediate width lines of post-shocked hydrogen and helium dominate the spectrum. We therefore focused on using these intermediate width lines to diagnose dust formation in SN~2005ip at later epochs. 

The intermediate-width He\,{\sc i}~7065\,\AA\ line profile that has become prominent in the optical spectrum by $\sim$413\,d  appears initially symmetrical before becoming increasingly asymmetrical, with a flux bias towards the blue (see Figure \ref{fig_7065}). It has been suggested that the increasing attenuation of the red side of the line profile at late times ($t>400$~d), indicative of dust formation, could be the result of dust formation within the post-shock region \citep{Smith2009,Fox2010,Smith2017}, with asymmetries in the broad lines invoked as evidence for ejecta dust formation (for example Figure \ref{fig_graphic_late_ps_dust}).  However, we consider here the possibility that dust formation in the ejecta alone could account for the blueshifting observed in the intermediate width profiles. In a shellular model, the ejecta dust is located interior to the post-shock gas and could theoretically induce the observed asymmetry (see Figure \ref{fig_graphic_late_ej_dust}). In our models, we seek to understand whether ejecta-condensed dust formation in SN~2005ip could account for these blueshifted intermediate width lines and to quantify the masses of dust that would be required in order to reproduce the observed H$\alpha$, H$\beta$ and He~{\sc i} line profiles. 

\subsubsection{Model Details}
\label{scn_int_model_details}
We have modelled the intermediate-width ($\sim2000$\,km\,s$^{-1}$) optical line profiles of SN~2005ip at a range of epochs, namely 413\,d, 905\,d, 2242\,d, 3099\,d and 4075\,d.  At epochs 413\,d and 905\,d, only He\,{\sc i}\,7065\,\AA\ is modelled since the intermediate-width H$\alpha$ profile that arises in the post-shock region is complicated by a significant contribution from the broad H$\alpha$ line from the ejecta.  For the last three epochs, however, three lines were modelled simultaneously. At 2242\,d and 3099\,d, H$\alpha$, H$\beta$ and He~{\sc i}\,7065\,\AA\ were modelled. Since the spectrum extended further into the IR at 4075\,d, we were able to instead model H$\alpha$, H$\beta$ and He~{\sc i}\,10830\,\AA\ at this epoch.
We  consider the geometry described by \citet{Smith2009} with an outward blast wave expanding at $\sim$14,000\,km\,s$^{-1}$ through a clumpy, shell-like CSM.  The blast wave shocks the clumps that it passes, exciting and accelerating the clumps from their initial velocity at $\sim$120\,km\,s$^{-1}$ imparted from the progenitor's stellar wind to as much as $\sim2000$\,km\,s$^{-1}$. The post-shock region comprises CSM shocked by the blast wave and ejecta shocked by the reverse shock, which we assume to be mixed together in a shell surrounding the unshocked expanding ejecta.  

The velocity distribution in this region is more complex than the freely-expanding ejecta and is generally much slower ($\sim$2000~km\,s$^{-1}$). In our models, we allowed the emitting gas to follow a velocity distribution $\propto v^{-\alpha_{\rm vel}}$ between a lower bound, $v_{\rm min}$, and an upper bound, $v_{\rm max}$, independent of radius.  The line emitting region, in this case representing the post-shocked gas, was a shell bounded by an outer radius and an inner radius defined respectively as
\begin{equation}
\begin{aligned}
\label{eqn_radii_out}
\frac{R_{\rm out}}{\rm cm} &= 8.64 \times 10^{9} \times  \frac{14,000}{{\rm \,km\,s}^{-1}} \times \frac{t}{{\rm days}} &&\\
&= 1.2096 \times 10^{14} \times \frac{t}{{\rm days}}
\end{aligned}
\end{equation}
\begin{equation}
\begin{aligned}
\label{eqn_radii_in}
\frac{R_{\rm in}}{\rm cm} &= \frac{R_{\rm min}}{\rm cm} + 8.64 \times 10^{9}  \left( \frac{300}{\rm km\,s^{-1}} \times \frac{(t-413)}{\rm days} \right) \\
&= 5\times 10^{16} + 2.592 \times 10^{12} \times \frac{(t-413)}{\rm days},
\end{aligned}
\end{equation}

\noindent where $t$ is the number of days post-discovery and $R_{\rm min}$ is the minimum radius of the CSM at an estimated time of initial impact between the blast wave and the CSM (taken to be $R_{\rm min}=5 \times 10^{16}$cm based on a blast wave traveling at 14,000\,km\,s$^{-1}$ impacting dense CSM at $\sim$413~d when the intermediate width He\,{\sc i}~7065~\AA\ line first appears).  We assume that the forward shock continues to propagate at a speed of approximately 14,000~km\,s$^{-1}$ as it progresses through the gaps in the extended, clumpy CSM \citep{Smith2009}.  300~km\,s$^{-1}$ is adopted as the rate of advancement of the inner radius of the post-shock region based on the minimum velocity of the  material in this region as derived from the best-fitting dust free He\,{\sc i}~7065~\AA\ model at 413~d. The radii of the line emitting post-shock region are given in Table \ref{tb_later_time_radii}.

Despite our primary interest being in the mass and properties of the ejecta dust treated in the models, the spatial extent of the line emitting region must be considered since the relative volumes of the line emitting region and the dusty ejecta interior to it determine the extent to which the radiation is occulted. The smaller the dusty region interior to the emitting region, the more opportunity the radiation has to reach us unattenuated.

	We note that this model does not take into account the passage of the reverse shock into the ejecta. However, the ratio of the reverse shock position to that of the contact surface between the ejecta and the CSM is very large ($\sim$0.98), particularly so for the strongly radiative regime \citep{Chevalier1982}. The evolution of $R_{\rm in}$ is therefore taken as an approximation to the radius of the reverse shock.  This is consistent with the approach adopted by \citet{Chugai2018} for the similar object SN~2010jl. 

The emissivity distribution is proportional to the square of the gas density distribution, which follows a smooth power-law $\propto r^{-\beta_{\rm gas}}$ between $R_{\rm in}$ and $R_{\rm out}$. Whilst it is anticipated that the gas is located in dense CSM clumps, there was little difference between smooth and clumped gas density distributions and smooth models ran significantly faster. 

For the gas density and velocity distribution described above, we considered the dust to be located in the ejecta,  entirely interior to the post-shock, line emitting shell as illustrated in Figure \ref{fig_graphic_late_ej_dust}.  The outer radius of the dusty ejecta ($R_{\rm out(dust)}$) is taken to be equal to the inner radius of the post-shock region ($R_{\rm in}$) for continuity. The inner radius of the dusty shell ($R_{\rm in(dust)}$) was fixed at $0.01R_{\rm out(dust)}$. Initial manual investigations indicated that inferred properties of the models were not very sensitive to changes in the value of $R_{\rm in(dust)}$. Dust was entirely contained in clumps of size $R_{\rm out}/25$ stochastically distributed according to a power-law $\propto r^{-2}$ and located interior to the post-shock region ($<R_{\rm in}$). At the earliest epochs, therefore, clumps were $<R_{\rm in}/10$ and at the latest epoch (4075\,d) clumps were $\sim R_{\rm in}/3$. The inner radius of the dust shell was $0.01R_{\rm in}$ in all cases. We do not rule out the possibility of dust formation in the post-shock region ($R_{\rm in}<r<R_{\rm out}$) as illustrated in Figure \ref{fig_graphic_late_ps_dust}, but models of this scenario were beyond the scope of this paper. 

\begin{figure}
\centering
\includegraphics[width = 0.8\linewidth]{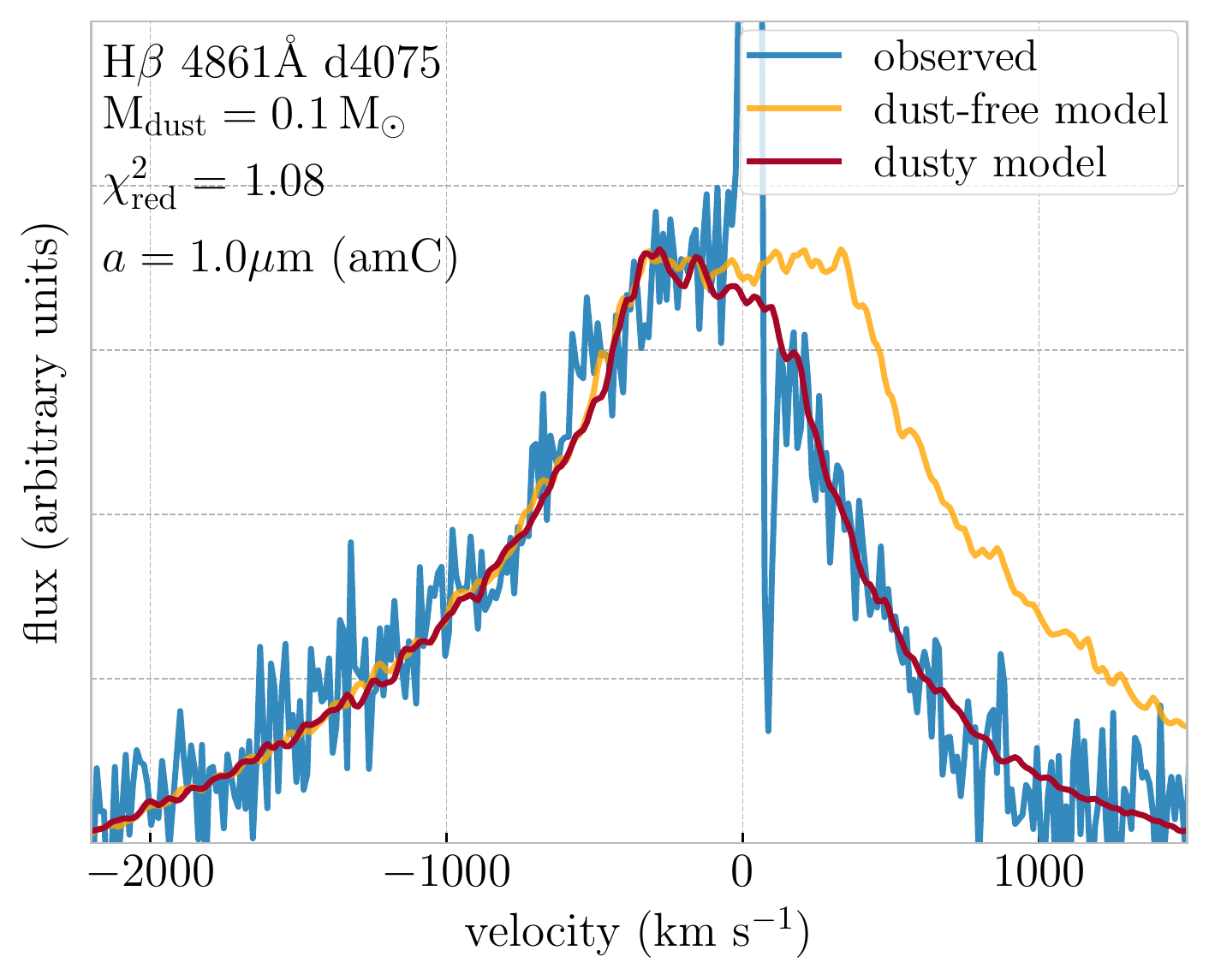}

\includegraphics[width = 0.85\linewidth,trim = -10 0 0 0]{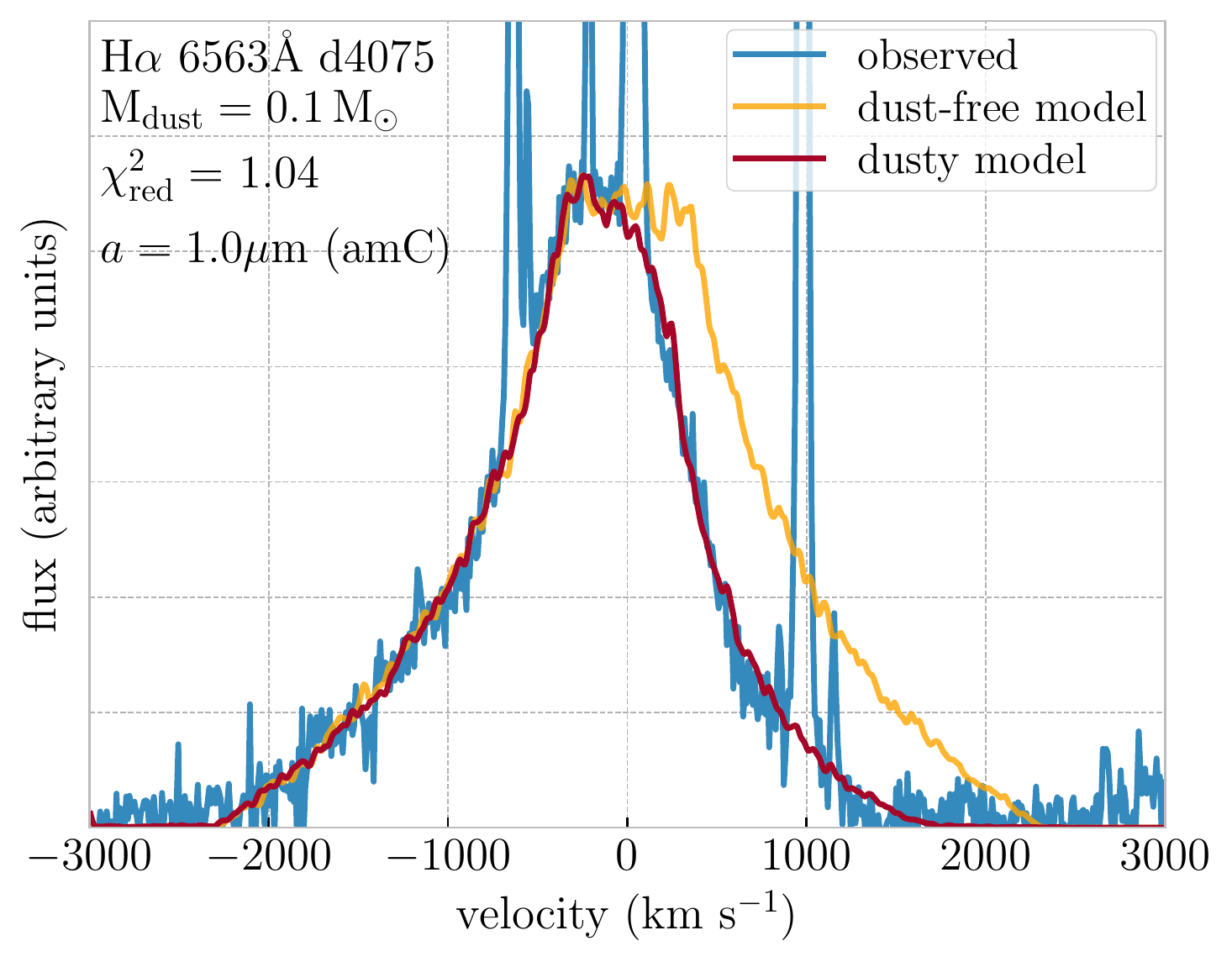}

\includegraphics[width = 0.8\linewidth]{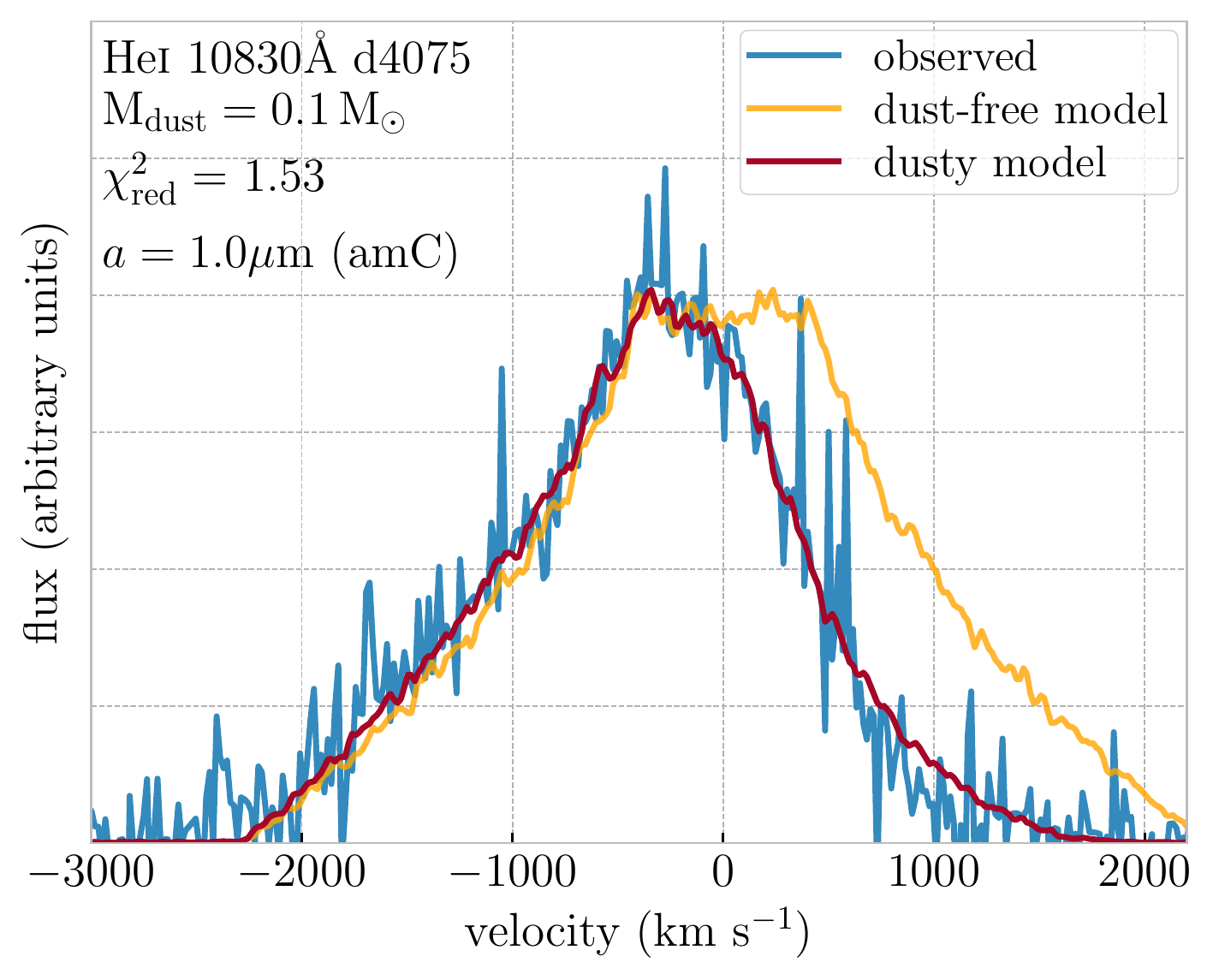}
\caption{An example of consistent set of best-fitting line profiles obtained manually for the intermediate width H$\alpha$. H$\beta$ and  line at 4075~d post-discovery.}
\label{fig_late_time_best_fits}
\end{figure}

An initial manual investigation of parameter space was performed in order to understand the appropriateness of each model and the prior range that should be investigated for each parameter using the automated MCMC routine. An example of a set of best-fits obtained manually for the H$\alpha$, H$\beta$ and He~{\sc i}~7065~\AA\ lines at d4075 is presented in Figure \ref{fig_late_time_best_fits}.

In all cases, seven variable parameters  (as described above) were investigated. These are summarised, along with the adopted priors, in Table \ref{tb_priors}. The radii defining the post-shock region at later epochs are described in Table \ref{tb_later_time_radii}.
\begin{table}
	\caption{Outer and inner radii of the line emitting post-shock region as defined respectively by Equations  \ref{eqn_radii_out} and \ref{eqn_radii_in}.  Dust is contained entirely within the ejecta which extends to the inner edge of the post-shock region.}
	\label{tb_later_time_radii}
	\centering
	\begin{tabular*}{\linewidth}{C{0.25\linewidth} C{0.25\linewidth} C{0.25\linewidth}}
		\hline
		Epoch  & $R_{\rm out}$  & $R_{\rm in}$ \\
		(d)  & ($10^{17}$\,cm)  & ($10^{17}$\,cm) \\
		\hline
		413 &0.50	&	0.50\\
		905 &1.1&	0.51\\ 
		2242 &2.7	&	0.55\\ 
		3099 &3.8	&	0.57\\ 
		4075 &4.9	&	0.59 \\ 
		\hline
	\end{tabular*}
\end{table}

\subsubsection{Ejecta dust formation at late times}

\begin{figure*}
\includegraphics[clip = true, trim = 0 0 25 0, width = \linewidth]{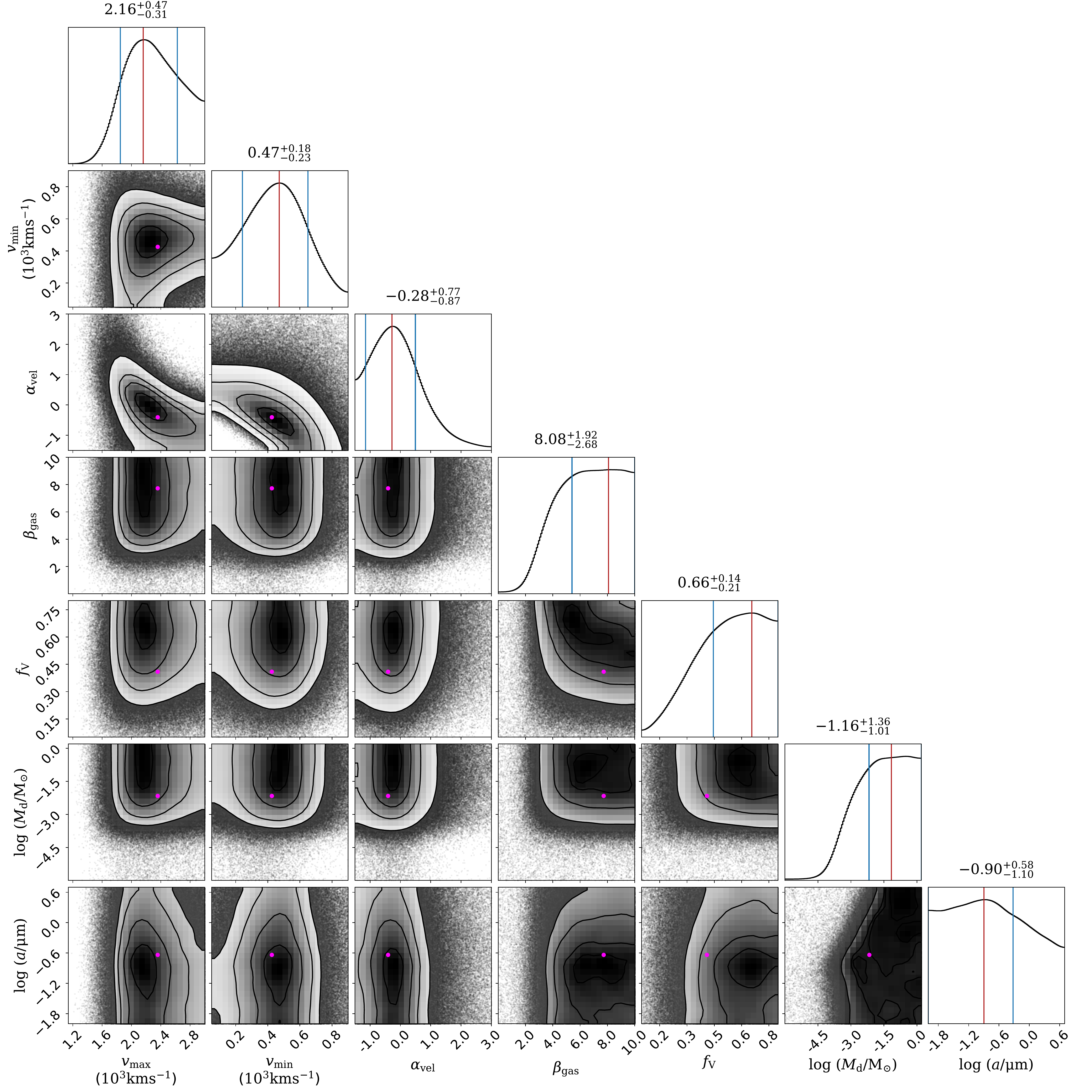}
\caption{The full 7D posterior distribution for the MCMC model of the intermediate width H$\alpha$, H$\beta$ and He\,{\sc i}\,10830\,\AA\ lines of SN~2005ip at 4075~d. The vertical red lines indicates the mode whilst the blue vertical lines indicate the bounds of the 68\% credibility region. The pink circle indicates the best-fitting parameter set from the MCMC run.}
\label{fig_d4075_full_posterior}
\end{figure*}

\renewcommand{\arraystretch}{1.6}
	\renewcommand{\tabcolsep}{0.32cm}
\begin{table}
	\caption{Dust mass and grain radius estimates for the amorphous carbon models of the intermediate width optical and near-IR emission lines of SN~2005ip at late times ($t>400$~d). The modal value from the marginalised 1D dust mass and grain radius distribution is given in addition to the 68\% confidence interval. A $^*$ indicates that the upper limit is equal to the upper bound of the prior range. }
	\label{tb_dust_masses_late}
    \centering
  	\begin{tabular*}{\linewidth}{C{0.4\linewidth} C{0.4\linewidth} }
	\hline
	Epoch  & $\log M_{\rm d}$  \\
	(d)  & ($\log$ M$_{\odot}$)\\
	\hline
	413 &$ -5.14 \substack{ +0.61 \\ -0.86 }$\\
	905 &$ -2.72 \substack{ +1.11 \\ -1.37 }$\\ 
	2242 &$ -1.06 \substack{ +1.26 \\ -0.60 ^*}$\\ 
	3099 &$ -1.11 \substack{ +1.31 \\ -0.92 ^*}$\\ 
	4075 &$ -1.16 \substack{ +1.36 \\ -1.01 ^*}$\\ 
	\hline
\end{tabular*}
\end{table}
\begin{figure*}
\includegraphics[clip = true, trim = 0 0 0 0, width = 0.2\linewidth]{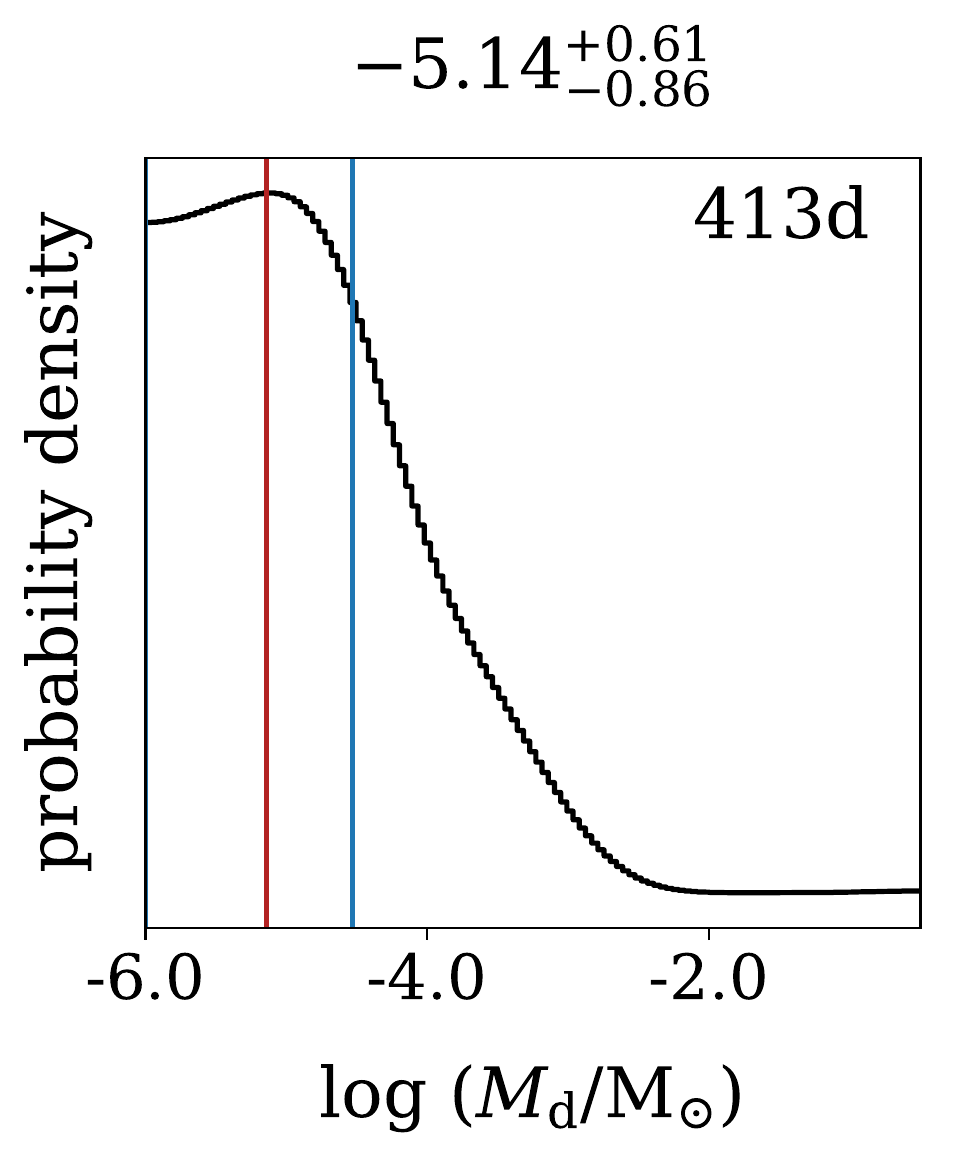}
\includegraphics[clip = true, trim = 27 0 0 0, width = 0.18\linewidth]{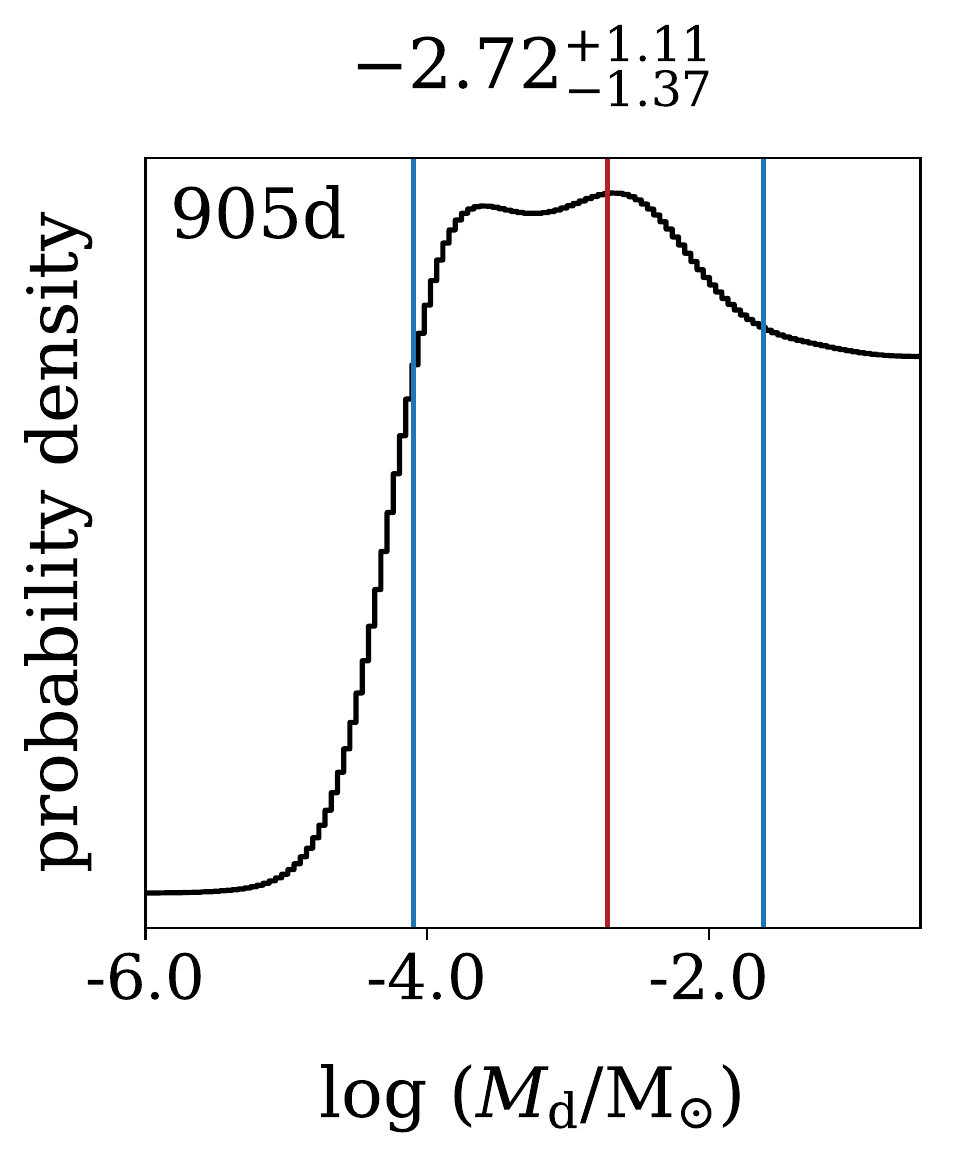}
\includegraphics[clip = true, trim = 27 0 0 0, width = 0.18\linewidth]{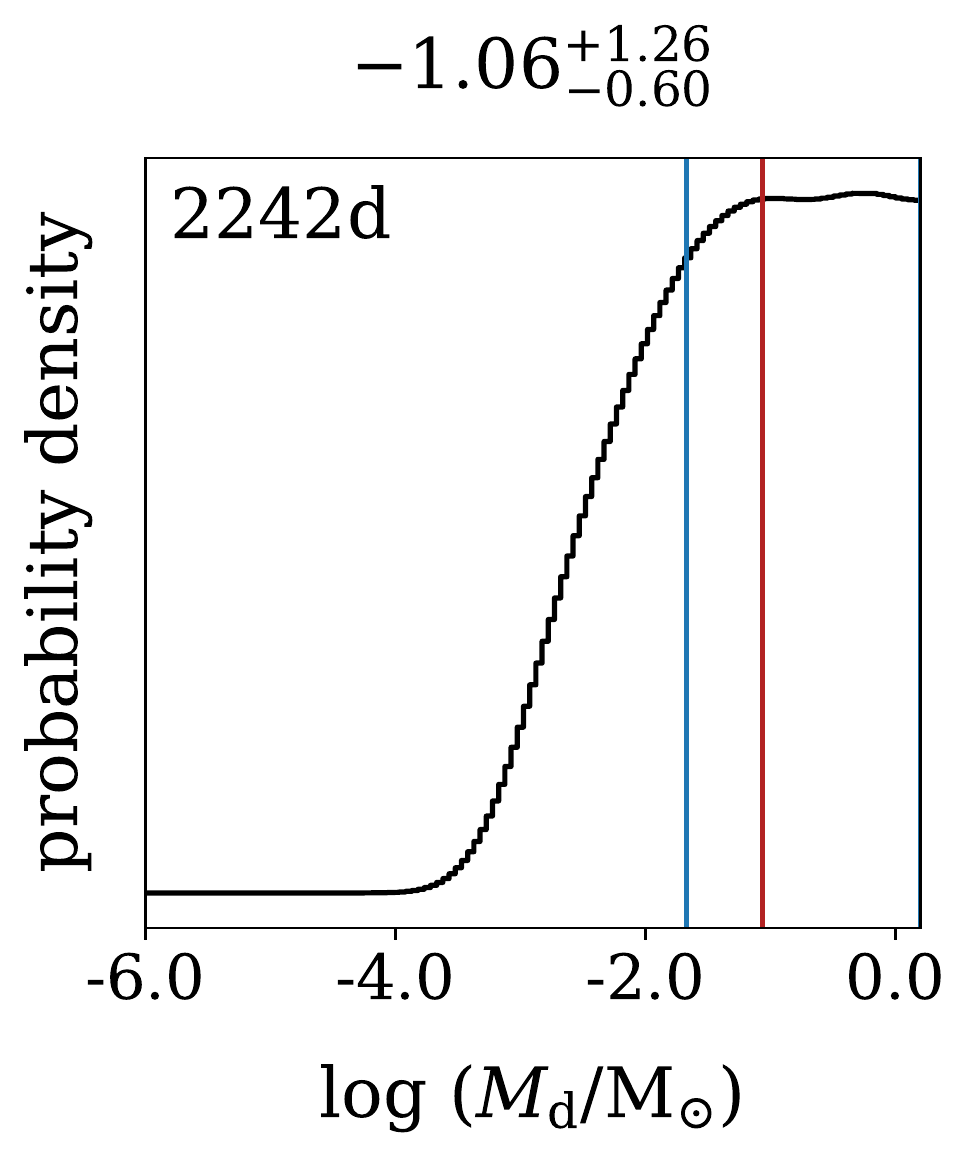}
\includegraphics[clip = true, trim = 27 0 0 0, width = 0.18\linewidth]{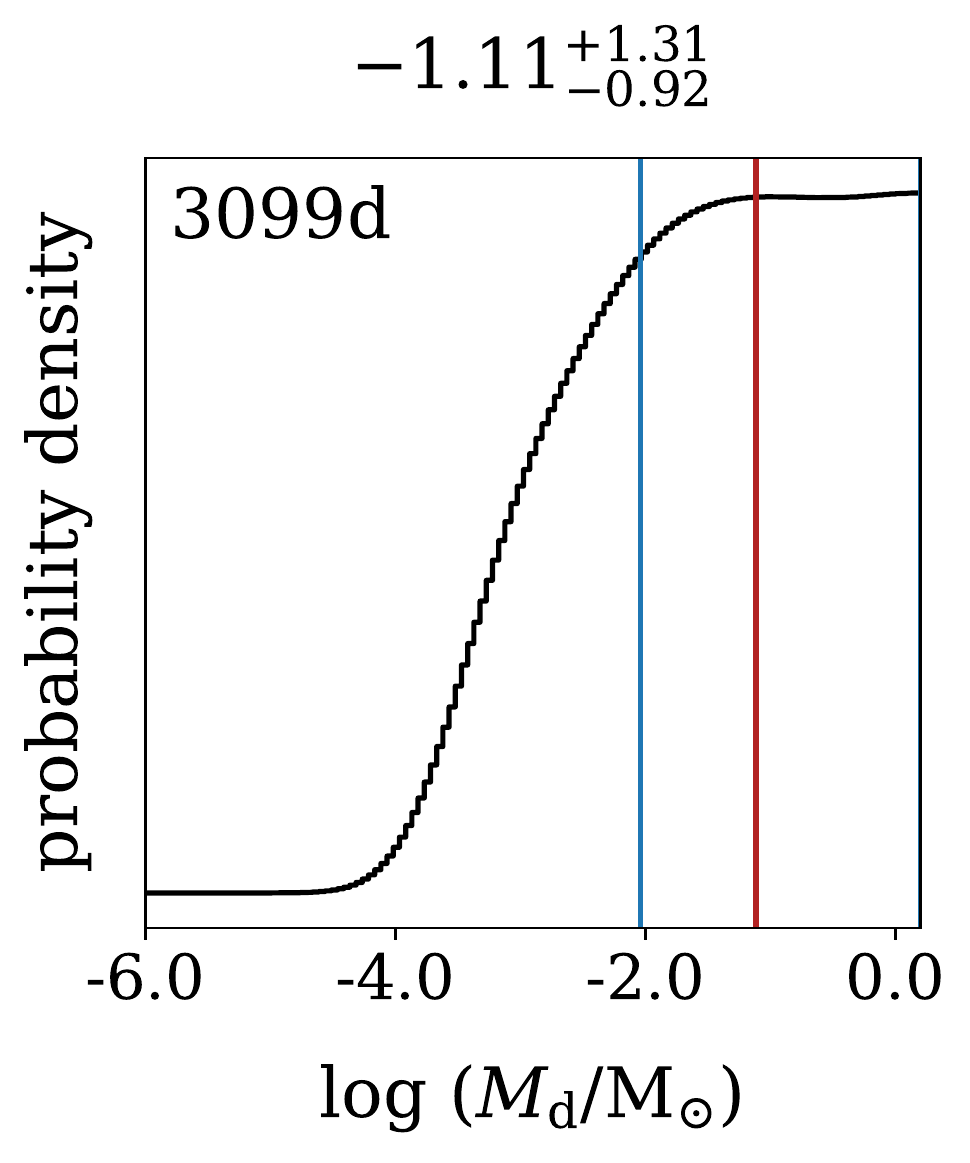}
\includegraphics[clip = true, trim = 27 0 0 0, width = 0.18\linewidth]{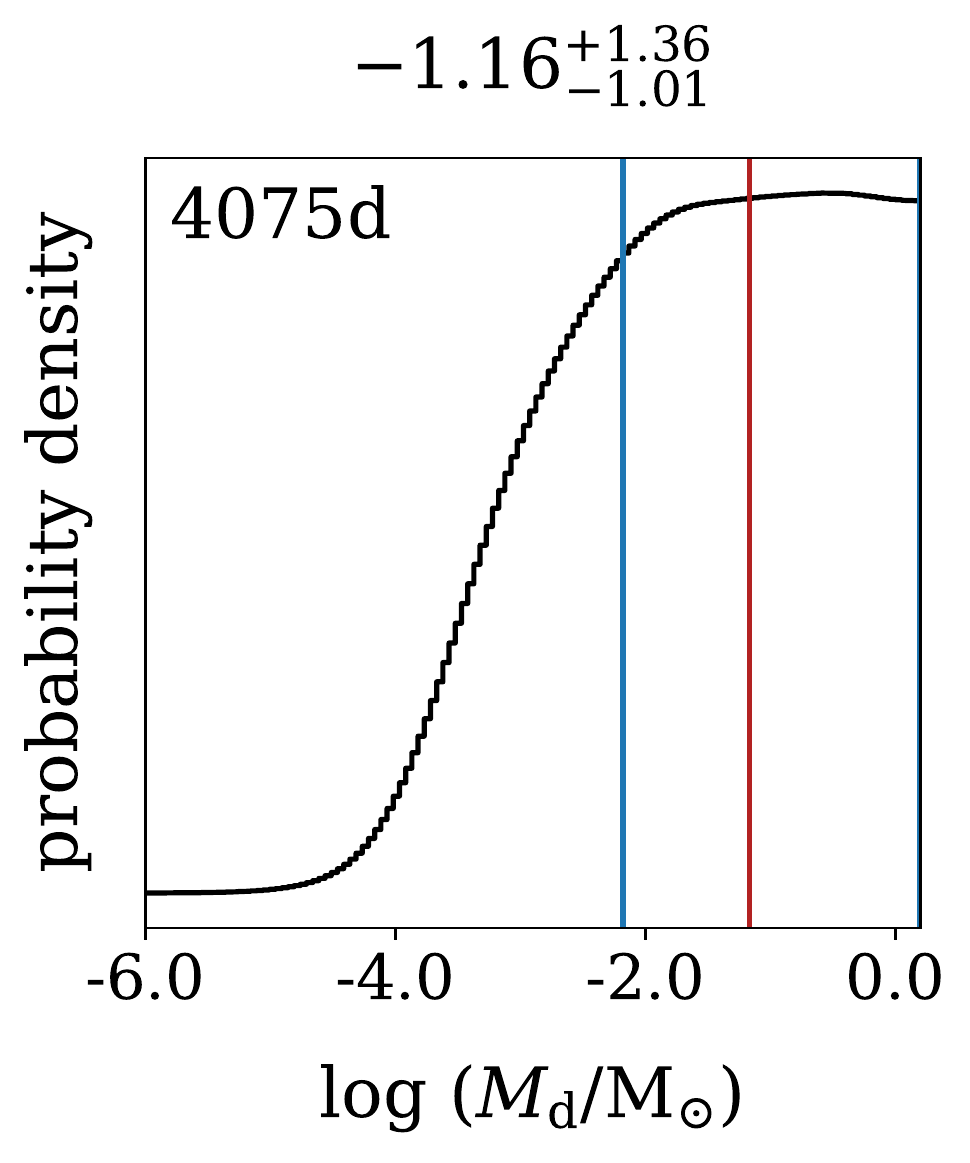}
\caption{Dust mass 1D marginalised posterior probability distribution for the later-time models. Vertical blue lines indicate the upper and lower bounds of the shortest region containing 68\% of the probability density (i.e. the region of highest probability density). Red vertical lines indicate the modal dust mass. The upper limits indicated for the latter three epochs are equal to the upper limit of the prior range.}
\label{fig_late_1D_dustmass}
\end{figure*}
The full 7D posteriors for all epochs are presented in the online material. We include the full posterior for the 4075~d model here for illustration purposes (Figure \ref{fig_d4075_full_posterior}). The best-fitting set of parameters from the MCMC run is marked on the contour plots for comparison. The modal parameter value along with the 68\% credibility region for each 1D marginalised distribution for all parameters and epochs is presented in full in the Appendix (Table \ref{tb_results_full}) and a summary of the dust masses are presented in Table \ref{tb_dust_masses_late}. 

It can be seen from this posterior (and it is the case for the other epochs as well), that the majority of parameters do not correlate with each other. There are, however, some predictable dependencies between, for example, the dust mass and dust grain radius. Since the wavelength dependence of dust extinction varies with grain radius, the grain radius can be significantly better constrained by modelling multiple lines across the spectrum simultaneously than with a single line. A broad peak in the grain radius 1D marginalised probability distribution is seen at $\sim0.1${\si{\um}}, approximately the grain radius at which amorphous carbon has the highest opacity, at all epochs with the exception of the symmetric He\,{\sc i}\,7065\,\AA\ line at 413~d. At 413~d an aritifical preference for larger grains, which are less absorbing, is shown for the same reasons described in Section \ref{early_model_results}. This should not be interpreted as an indication that larger grains are more likely at this epoch (see Section \ref{scn_discussion} for a wider interpretation of these results).

Importantly, however, by investigating a reasonable range of single  grain radii with an appropriate prior, we can understand the likelihood distribution of the ejecta dust mass in SN~2005ip marginalised over the grain radius of the dust. The 1D marginalised probability functions for the dust mass at the later epochs are shown in Figure \ref{fig_late_1D_dustmass}. The symmetry of the He~{\sc i} profile at 413~d limits the amount of dust that can be present at this time. Consequently, a sharp decline in the likelihood of dust masses higher than $\sim 3 \times 10^{-5}$~M$_{\odot}$ is seen for this epoch.

	An important factor in determining the the location of the dust formation in SN~2005ip is whether the ejecta dust mass that causes the observed asymmetries in the broad line profiles at early epochs is consistent with the symmetric intermediate width He\,{\sc i} line profile seen at 413~d.  The model at 169\,d requires a dust mass of $\sim 7 \times 10^{-6}$~M$_{\odot}$. This dust mass would not discernibly affect the intermediate width line profile at 413\,d and is therefore consistent with the symmetric intermediate width He~{\sc i} line at 413\,d . The uncertainties in the inferred dust masses at 169\,d and 413\,d make in unclear whether there is continued dust mass growth during this period or whether there is a plateau in the formation rate, which might be suggestive of two distinct periods of dust formation.

As the intermediate width profiles become increasingly asymmetric, the shape of the dust mass probability density distribution changes. A sharp decline is now seen in the distribution for smaller masses of dust and we can infer that by 905~d, $\sim10^{-3}$~M$_{\odot}$ of dust has formed. This appears to be the time when dust is forming fastest in the ejecta. By 2242~d, the dust formation rate is beginning to slow but is still increasing.  By 4075\,d, the models require that more than $10^{-2}$~M$_{\odot}$ of ejecta-condensed dust has formed in SN~2005ip, with a most likely dust mass of $\sim$0.1~M$_{\odot}$. We note that the plateau in the dust mass posterior distribution at the later epochs is likely the result of optically thick dusty clumps that can ``hide'' dust within them. It is therefore not possible to place an upper limit on the dust mass that has formed by 2242\,d, 3099\,d and 4075\,d from these models. At earlier epochs, however, the dust mass is constrained.

Whilst there is no dependence at all on the volume filling factor occupied by dust clumps ($f_{\rm V}$) at 413~d, at later epochs there is a skew in the posterior toward higher filling factors. In general, a minimum dust volume filling factor of $\sim$0.6 is required in order to reproduce the observed profiles. 

The other parameters are  all relatively well constrained. The posterior distribution of the parameter $\beta_{\rm gas}$, which controls the steepness of the post-shock emissivity distribution between $R_{\rm in}$ and $R_{\rm out}$, indicates that the density distribution of the post-shock gas is more concentrated towards the inner radius, as would be expected. Both maximum and minimum velocities are well constrained at all epochs.

\section{Discussion}
\label{scn_discussion}

\subsection{Dust formation in SN~2005ip at early times}
Dust formation associated with SN~2005ip has been inferred in previous works from the appearance and persistence of both an excess IR flux and blueshifted optical and near-IR line profiles \citep{Fox2009,Smith2009,Fox2009,Fox2010,Stritzinger2012,Smith2017}. During the first 200~d post-discovery, the broad H$\alpha$ line profile became shifted towards the blue. This was attributed to dust formation in the ejecta. The simultaneous appearance of an IR excess from hot dust grains was presumed to be associated with this newly-formed ejecta condensed dust \citep{Smith2009}. 

We find that the early time broad H$\alpha$ profile can be fitted very well with a relatively simple model of clumped ejecta dust punctuating a smooth distribution of gas. Only very small masses of dust are required to produce the asymmetries observed at these very early times since the radius of the ejecta is still small. We determine that a dust mass of $\sim$10$^{-5}$\,M$_{\odot}$ has condensed in the ejecta by 169~d. Recent analysis by  \citet{Nielsen2018} identified an acceleration in the decline of the optical light curve commencing at around 50 days. Dust formation in the ejecta would explain this sudden drop. By comparing the light curve to theoretical curves, \citet{Nielsen2018} estimated that, by 150~d, an additional extinction of approximately $A_{\rm V}=1$ mag would be required to account for the drop in optical luminosity. Our best-fitting model at 169~d as derived from the MCMC run is representative of the most likely regions of the parameter space (this is not always necessarily the case). This model has an average optical depth between $R_{\rm in}$ and $R_{\rm out}$ in the V band  of 0.84, which is consistent with the value derived by \citet{Nielsen2018}.

The maximum radial velocity of the line is also well constrained by the models and suggests a modal maximum velocity of $\sim$12,000~km~s$^{-1}$ at 169\,d post-outburst. Under simple homologous expansion, this yields an outer radius of the ejecta of 1.8 $\times 10^{16}$~cm which is consistent with the minimum blackbody radius of $\sim10^{16}$cm at 200\,d derived by \citet{Fox2009}.

\subsection{Dust formation in SN~2005ip at later times}
At later times, the attenuation of the red side of the intermediate width line profiles has been attributed to dust formation in the post-shock region and SN~2005ip has been cited as one of the first supernovae for which there is strong evidence of dust formation in {\em both} the ejecta and the post-shock region \citep{Smith2009}. Whilst we do not preclude dust formation in the post-shock region, we have shown that it is not required to account for the blueshifting observed in the intermediate width line profiles that arise in the post-shock gas. Continued formation of dust in the ejecta is perhaps a simpler explanation for the blueshifted intermediate width line profiles.

\renewcommand{\arraystretch}{1.6}
	\renewcommand{\tabcolsep}{0.1cm}
\begin{table}
	\centering
    \caption{The 936\,d warm (400\,K) and hot (800\,K) dust mass estimates derived by \citet{Fox2010} (F10 here)  from {\em Spitzer} observations for two different grain sizes with their associated blackbody radii. Our approximate dust mass estimates are given for equivalent dust grain radii along with the radius adopted for the ejecta dust distribution at 905\,d.}
	\begin{tabular}{L{2.1cm} C{1.8cm} C{1.8cm} C{1.8cm}}
		Dust component & $M_{\rm dust}$ ($a = 0.1 \mu$m)   (M$_{\odot}$) & $M_{\rm dust}$ ($a = 1.0 \mu$m) (M$_{\odot}$) &  Radius of dust shell (cm) \\
		\hline
			 Hot, near-IR dust (F10) & $\sim5 \times 10^{-4}$ & - & $>7.7 \times 10^{15}$\\
			 Warm, mid-IR dust (F10) & $\sim4 \times 10^{-2}$ & $\sim5 \times 10^{-3}$ & $>4.8 \times 10^{16}$ \\
			 
			 \hline
			 Ejecta dust (this work) & $\sim2 \times 10^{-4}$ & $\sim2 \times 10^{-3}$ & $5.1 \times 10^{16}$\\
		\hline
		
	\end{tabular}
\label{tb_dust_masses_radii}
\end{table}

Analysis of IR observations by both \citet{Fox2010} and \citet{Stritzinger2012} has suggested that the IR flux has at least two different dust temperature components, one a cooler component associated with the SN itself, and another a warmer component that is likely associated with pre-existing warm dust that has been heated by optical and X-ray emission from the interaction of the blast wave with the dense CSM \citep[also][]{Fox2011}.  Based on near-IR observations and mid-IR observations at 936~d using the {\em Spitzer} Infrared Spectrograph and 3.6{\si{\um}} and 4.5{\si{\um}} bands, \citet{Fox2010} derived a warm ($\sim$400\,K) dust mass of $\sim$4$\times 10^{-2}$\,M$_{\odot}$ and a hot ($\sim$800\,K) dust mass of $\sim$5$\times 10^{-4}$\,M$_{\odot}$ for graphite grains with radii $\sim 0.1${\si{\um}}, or alternatively a warm dust mass of $\sim$5$\times 10^{-3}$\,M$_{\odot}$ for grains of radius $\sim$1.0\,\si{\um}. We summarise these dust masses in Table \ref{tb_dust_masses_radii} along with their respective corresponding blackbody radii of $4.8 \times 10^{16}$cm and $7.7 \times 10^{15}$cm. Based on our model of the He\,{\sc i}\,7065\,\AA\ line at 905~d and the explicit 2D marginalised posterior for dust mass and grain radius, we deduce a dust masses of $\sim$2$\times 10^{-4}$\,M$_{\odot}$ and $\sim$2$\times 10^{-3}$\,M$_{\odot}$ for grains of radii 0.1\,{\si{um}} and 1.0\,{\si{um}} respectively. The inner radius of the post-shock region, and therefore outer radius of the dusty ejecta, was $5.1 \times 10^{16}$cm (as per Equation \ref{eqn_radii_in}). These values are included in Table \ref{tb_dust_masses_radii} for ease of comparison with \citet{Fox2010}.

We find that for dust grains of 1.0\,{\si{um}}, our dust mass estimate of $\sim$2$\times 10^{-3}$\,M$_{\odot}$ is consistent with the warm dust component identified by \citet{Fox2010}, suggesting that its origin could be in newly-formed dust in the ejecta, not in pre-existing dust as suggested by these authors. However, for smaller grains of  radius 0.1\,{\si{\um}}, our dust mass estimate of $\sim$2$\times 10^{-4}$\,M$_{\odot}$ is consistent with the hot, near-IR component identified by \citet{Fox2010}, and would not account for the larger mass of warm, mid-IR dust. This scenario is generally consistent with that proposed by \citet{Fox2010}  wherein the hot dust component is associated with newly-formed dust, either in the ejecta (as here) or the post-shock region, and the warm dust component is associated with pre-existing dust in a distant circumstellar shell. 

At later times ($\sim$1500~d), however, the dust mass required to account for the hot dust component does not appear to significantly increase in mass \citep{Stritzinger2012}, whilst our models require an increasingly large dust mass to account for the observed red-blue asymmetries. It is possible that, as in SN~1987A, this larger dust mass is significantly colder than the warmer dust ($>400$\,K) detected by {\em Spitzer} and is therefore not accounted for in their analyses.

We note that the passage of the reverse shock is expected to destroy some of the dust, though the fraction of dust destroyed depends highly on the composition, grain size distribution and density distribution of the dust. The increasing dust mass inferred from our models is not necessarily representative of the dust formation rate since some of the dust in the outer regions of the ejecta may be have been destroyed by the passage of the reverse shock. An increasing dust mass over these epochs can be reasonably explained by requiring that the dust mass growth rate exceeds the rate of destruction.

\subsection{A comparison with SN 1987A}
\label{scn_discuss_87A_comparison}

\begin{figure}
\centering
\includegraphics[clip = true, trim = 0 0 0 0, width=\linewidth]{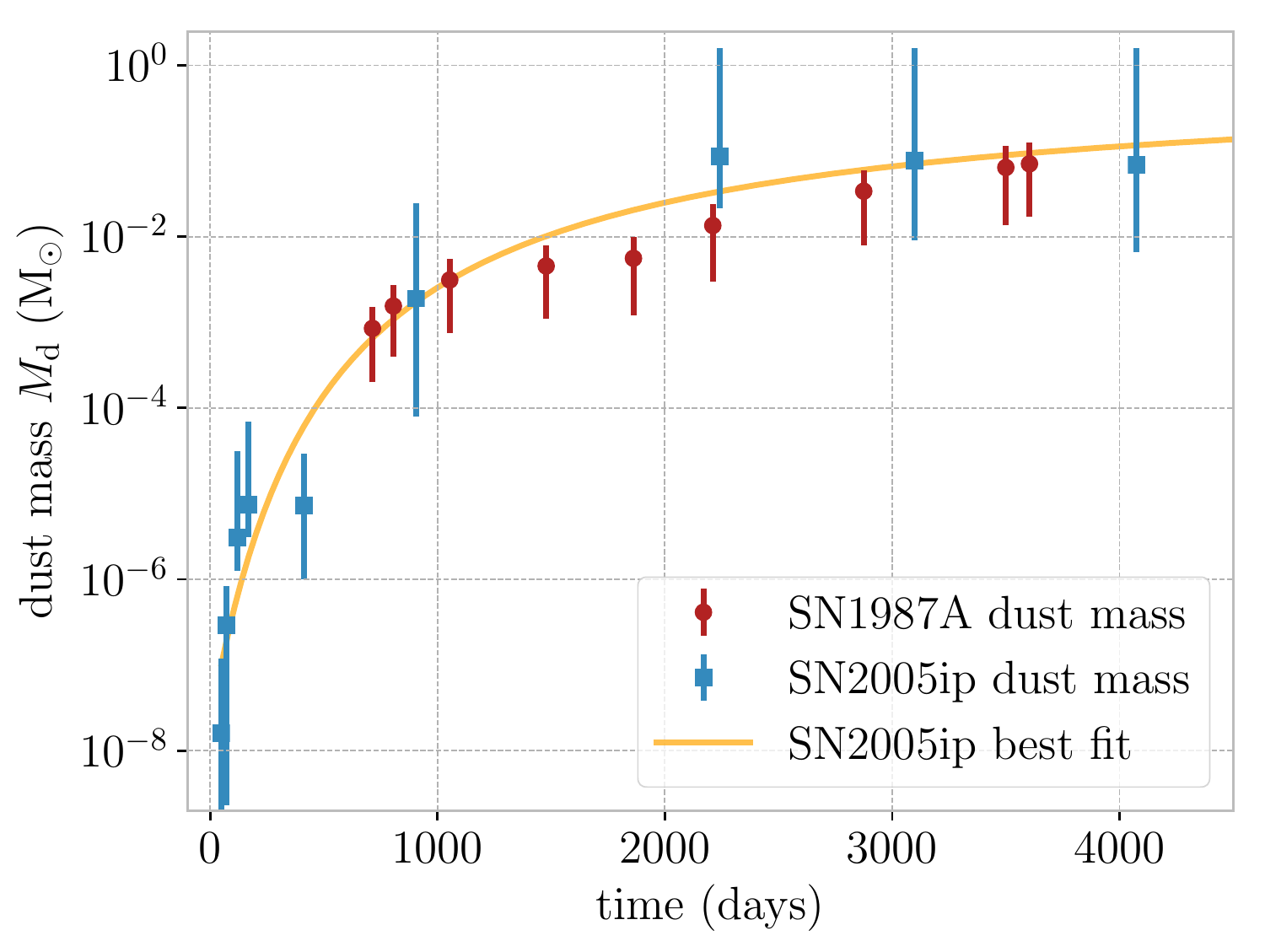}
\caption{The evolution of the ejecta dust mass in SN2005ip from line profile models ({\em blue}) with the best-fitting curve of the form $-\log(M_{\rm dust}/M_{\odot}) = c(a\frac{t}{\rm days}+b)^{-1}+d$ ({\em yellow}). Also plotted is the evolution of the dust mass in SN~1987A as determined from line profile models \citep[{\em red};][]{Bevan2016}.}
\label{fig_dust_mass_evol}
\end{figure}

We have shown that the observed asymmetries in both the broad and intermediate width optical and near-IR line profiles can be explained by dust formation in the ejecta of SN~2005ip. We have modelled this scenario and quantified the dust mass that would be required to reproduce the observations. We summarise the dust mass evolution in Figure \ref{fig_dust_mass_evol} and also present the best-fitting curve of the form $-\log(M_{\rm dust}/M_{\odot}) = c(a\frac{t}{\rm days}+b)^{-1}+d$. The values required are $a=0.0076,b=3.66,c=27.7$, and $d=-0.13$, where d represents the final log dust mass should no subsequent destruction of dust take place.

We initially consider the comparison between the evolution of dust in SN~2005ip with the time evolution of dust in SN~1987A, the only other object for which the long term ($>10$\,yr) evolution of the formation of dust has been quantified. Detailed radiative transfer models of the optical and IR spectral energy distribution (SED) of SN~1987A have been shown to yield dust masses that are in strong agreement with those inferred from line profile fitting and deduce a dust mass today of the order of $\sim$0.5\,M$_{\odot}$ \citep{Matsuura2011,Matsuura2015,Wesson2015,Bevan2016}. Observations with the Atacama Large Millimetre Array (ALMA) have revealed that the dust responsible for the observed asymmetries in SN1987A's line profile and the late-time IR excess formed inside the ejecta \citep{Lucy1989,Indebetouw2014}. Whilst SN~1987A is a peculiar Type IIP SN that exploded in a very different environment to the IIn SN~2005ip, it is interesting to compare the time evolution of their dust formation. Given these intrinsic differences, it is particularly interesting to note the strong agreement between the dust masses inferred for SN~2005ip and SN~1987A at the same epochs (see Figure \ref{fig_dust_mass_evol}). Dust formation models  predict that supernova dust formation should have ceased within approximately 1000\,--\,1500 days of explosion \citep{Sarangi2015,Sluder2018}. The discovery of a continued steady increase in the dust mass after 1000 days in both of these supernovae  presents a challenge for these models. Further studies of the dust formation rate in other supernovae and supernovae remnants is required.

Like SN~1987A, there is no evidence of any silicate emission around 10{\si{\um}} \citep{Fox2010}, and so we have assumed that the dust associated with SN~2005ip is predominantly carbonaceous. This assumption is further supported by the similar IIN SN 2010jl which shows no evidence of any silicate dust features \citep{Williams2015}. However, the possibility of silicate dust should not be ruled out entirely. The lack of a 10{\si{\um}} feature could indicate cold silicate dust with an additional warm, featureless dust component or large silicate grains with $2\pi a > \lambda$ sufficient to wash out the feature. From previous models, such as those for SN~1987A, we know that silicate dust tends to increase the required dust mass by almost an order of magnitude.

Even for carbonaceous grains, however, the ejecta dust mass required to account for the optical and near-IR spectrum of SN~2005ip 11 years after outburst is $\sim$0.1\,M$_{\odot}$. This is similar not only to SN~1987A, but also to SN~1980K, SN~1993J and, at later times, Cas~A and the Crab Nebula \citep[e.g.][]{Owen2015,Bevan2017,deLooze2017} as well as a number of other Galactic and nearby supernova remnants \citep{Temim2017,Rho2018}. It is unclear whether the dust formation rate has peaked or whether, like in SN~1987A, the dust mass will continue to grow to several times its current mass. Future observations are required to continue monitoring the rate of dust formation (or destruction) in SN~2005ip.

Determining the contribution of core-collapse supernovae to the total dust budget of the early Universe depends critically on how much of the dust that forms in the ejecta of core-collapse supernovae can survive the passage of the reverse shock. In SN~2005ip, the reverse shock  has already arisen and will progress significantly more quickly than in most other CCSNe due to its dense CSM. It could therefore provide interesting insight into potential dust destruction in the future. 

\section{Conclusions}
\label{scn_conclusions}

We have presented a new late-time spectrum of SN~2005ip obtained with X-shooter on 2016 Dec 27 and 31. We also collated a number of other optical spectra from the archives of various telescopes in order to model the evolution of the optical and near-IR line profiles of SN~2005ip using a Bayesian approach. We have explored the possibility that dust formation in the ejecta of SN~2005ip could account for the observed blueshifting seen in both the broad H$\alpha$ line at early times and in the intermediate width Balmer series  and He\,{\sc i} lines at later times. We modelled this scenario using the line transfer code {\sc damocles} in combination with the Markov Chain Monte Carlo routine {\em emcee} for a range of epochs from 48~d to 4075~d post-discovery.

We conclude that, not only can dust formation in the ejecta of SN~2005ip account for the observed blueshifting in {\em both} the broad and intermediate width lines at all epochs, but that the inferred dust masses are remarkably consistent with the ejecta-condensed dust masses derived for SN~1987A. This is further evidence that core-collapse supernovae can form significant masses of dust in their ejecta and, if a significant fraction of their condensed dust masses survives subsequent reverse shock interactions, could potentially account for the large masses of dust seen in the early Universe. 

\section*{Acknowledgements}
This  work was supported by European Research Council (ERC) Advanced Grant SNDUST 694520. MM is supported by an STFC Ernest Rutherford fellowship (ST/L003597/1).

This paper is based on observations collected at the European Southern Observatory under ESO programme(s) 097.D-0525(A) and data obtained from the Weizmann Interactive Supernova Data Repository (WISeREP; https://wiserep.weizmann.ac.il), as well as publicly available spectra obtained from the archives of the observatories listed in Table 1. We thank the anonymous referee for their helpful suggestions.

\bibliography{sn2005ip_paper_clean}{}
\bibliographystyle{mnras}

\appendix
\section{Full results table}
The full results for each of the nine epochs modelled. The mode is given along with the 68\% confidence intervals. The accompanying plots of the posteriors are given in the online material.
\renewcommand{\arraystretch}{1.6}
\renewcommand{\tabcolsep}{0.32cm}
\begin{table*}
	\caption{Results for the modelled lines for each of the early epochs ({\em top}) and late epochs ({\em bottom}). $v_{\rm max}$ and $v_{\rm min}$ are respectively the maximum and minimum radial velocity of the emitting gas (which do not depend on radial position), $\beta_{\rm vel}$ is the index of the velocity power-law probability distribution, $\beta_{\rm gas}$ is the smooth gas density radial power-law distribution (on which the emissivity distribution depends), $\alpha_{\rm vel}$ is the index of the velocity power-law probability distribution, $f_V$ is the fraction of the ejecta volume occupied by dust clumps, $\log M_{\rm d}$ is the $\log$ of the dust mass. and  $\log a$ is the $\log$ of the single grain radius.  In some cases, the upper or lower limit derived from the fitting process is a product of the parameter range investigated i.e. equal to one bound of the prior range. We indicate where this is the case for the upper limit ($^*$) and the lower limit ($^\dagger$). Where the 1D marginalised posterior distribution is entirely flat, we do not present data.  These figures should be used in reference to the full 2D posteriors given in the online material.)}

	\label{tb_results_full}
	  	\begin{tabular*}{\linewidth}{C{1cm} C{1.8cm}C{1.8cm}C{1.6cm}C{1.6cm}C{1.6cm}C{1.6cm}C{1.6cm}l}
		\cline{1-8}
		Epoch  & $v_{\rm max}$  & $v_{\rm min}$ & $\beta_{\rm clump}$  & $\beta_{\rm gas}$ & $f_V$ & $\log M_{\rm d}$  & $\log a$ \\
		(d) & (10$^3$~km~s$^{-1}$)  & (10$^3$~km~s$^{-1}$) &   &  &  & ($\log$ M$_{\odot}$)  & ($\log$ \si{\um}) &\\
		\cline{1-8}
		48 &$ 16.71 \substack{ +1.40 \\ -1.08 }$
		     &$ 4.42 \substack{ +0.99 \\ -2.02 }$
		     &
		     &$ 0.85 \substack{ +0.33 \\ -0.28 }$
		     &
		     &$ -7.80 \substack{ +0.87 \\ -1.20 ^\dagger}$
		     &$ 0.69 \substack{ +0.01^*\\ -1.61 }$
		     &\\
		      
		72 &$ 15.14 \substack{ +2.00 \\ -1.31 }$
		     &$ 3.66 \substack{ +1.46 \\ -1.62 }$
		     &
		     &$ 0.91 \substack{ +0.39 \\ -0.42 }$
		     &
		     &$ -6.54 \substack{ +0.46 \\ -2.10 ^\dagger}$
		     &$ 0.49 \substack{ +0.21 ^* \\ -1.47 }$
		     &\\ 
		120 &$ 12.24 \substack{ +1.61 \\ -0.79 }$
		      &$ 2.70 \substack{ +1.00 \\ -1.61 }$
		      &$ 3.98 \substack{ +0.02^* \\ -2.36 }$
		      &$ 0.81 \substack{ +0.46 \\ -0.51 }$
		      &$ 0.05 \substack{ +0.47 \\ -0.00 ^\dagger}$
		      &$ -5.51 \substack{ +1.01 \\ -0.39 }$
		      &$ -1.21 \substack{ +1.08 \\ -0.70 }$
		      &\\ 
		169 &$ 11.70 \substack{ +1.59 \\ -1.03 }$
		      &$ 2.48 \substack{ +1.04 \\ -0.81 }$
		      &$ 3.98 \substack{ +0.02^* \\ -2.41 }$
		      &$ 1.03 \substack{ +0.48 \\ -0.51 }$
		      &$ 0.05 \substack{ +0.46 \\ -0.00^\dagger }$
		      &$ -5.13 \substack{ +0.97 \\ -0.38 }$ 
		      &$ -1.99 \substack{ +1.72 \\ -0.01^\dagger }$
		      &\\  
		\cline{1-8}
		\\
		\cline{1-8}
		Epoch  & $v_{\rm max}$  & $v_{\rm min}$ & $\alpha_{\rm vel}$  & $\beta_{\rm gas}$ & $f_V$ & $\log M_{\rm d}$  & $\log a$ \\
		(d) & (10$^3$~km~s$^{-1}$)  & (10$^3$~km~s$^{-1}$) &   &  &  & ($\log$ M$_{\odot}$)  & ($\log$ \si{\um}) &\\
		\cline{1-8}
		
		413 &$ 2.17 \substack{ +0.92 \\ -0.32 }$
		      &$ 0.55 \substack{ +0.21 \\ -0.19 }$
		      &$ -0.48 \substack{ +0.96 \\ -0.90 }$
		      &
		      &
		      &$ -5.14 \substack{ +0.61 \\ -0.86^\dagger }$
		      &$ 0.69 \substack{ +0.01 ^*\\ -1.57 }$
		      &\\ 
		905 &$ 1.66 \substack{ +0.39 \\ -0.29 }$
		       &$ 0.30 \substack{ +0.11 \\ -0.15 }$
		       &$ 0.04 \substack{ +0.56 \\ -0.91 }$
		       &$ 1.88 \substack{ +3.18 \\ -1.87 ^\dagger}$
		       &$ 0.05 \substack{ +0.47 \\ -0.00^\dagger }$
		       &$ -2.72 \substack{ +1.11 \\ -1.37 }$
		       &$ -0.97 \substack{ +0.67 \\ -1.02 }$
		       &\\ 
		2242 &$ 2.64 \substack{ +0.75 \\ -0.25 }$
		         &$ 0.73 \substack{ +0.06 \\ -0.06 }$
		         &$ -1.91 \substack{ +0.53 \\ -0.41 }$
		         &$ 9.97 \substack{ +0.03 ^*\\ -2.46 }$
		         &$ 0.66 \substack{ +0.12 \\ -0.09 }$
		         &$ -1.06 \substack{ +1.26 ^*\\ -0.60 }$
		         &$ -1.04 \substack{ +0.66 \\ -0.92 }$
		         &\\ 
		3099 &$ 2.33 \substack{ +0.23 \\ -0.17 }$
		         &$ 0.53 \substack{ +0.07 \\ -0.07 }$
		         &$ -1.09 \substack{ +0.40 \\ -0.35 }$
		         &$ 6.55 \substack{ +2.60 \\ -1.25 }$
		         &$ 0.59 \substack{ +0.14 \\ -0.13 }$
		         &$ -1.11 \substack{ +1.31 ^*\\ -0.92 }$
		         &$ -1.10 \substack{ +0.70 \\ -0.90^\dagger }$
		         &\\ 
		4075 &$ 2.16 \substack{ +0.47 \\ -0.31 }$
		         &$ 0.47 \substack{ +0.18 \\ -0.23 }$
		         &$ -0.28 \substack{ +0.77 \\ -0.87 }$
		         &$ 8.08 \substack{ +1.92^* \\ -2.68 }$
		         &$ 0.66 \substack{ +0.14^* \\ -0.21 }$
		         &$ -1.16 \substack{ +1.36^* \\ -1.01 }$
		         &$ -0.90 \substack{ +0.58 \\ -1.10^\dagger }$
		         &\\ 
		\cline{1-8}
	\end{tabular*}
\end{table*}

\label{lastpage}
\end{document}